%% file: ms.tex
\documentclass[preprint,journal]{vgtc}       

\usepackage{enumitem}
\usepackage{xspace}
\usepackage{changepage}

\include{customization}

\ifpdf
  \pdfoutput=1\relax                   
  \usepackage{graphicx}                
  \DeclareGraphicsExtensions{.pdf,.png,.jpg,.jpeg} 
\else
  \usepackage{graphicx}                
  \DeclareGraphicsExtensions{.eps}     
\fi%

\graphicspath{{figures/}{pictures/}{images/}{./}} 

\usepackage{microtype}                 
\PassOptionsToPackage{warn}{textcomp}  
\usepackage{textcomp}                  
\usepackage{mathptmx}                  
\usepackage{times}                     
\usepackage{cite}                      
\usepackage{tabu}                      
\usepackage{booktabs}                  

\ieeedoi{10.1109/TVCG.2020.3030460}

\onlineid{1243}

\vgtccategory{Research}
\vgtcpapertype{algorithm/technique}

\title{Personal Augmented Reality for Information Visualization\\ on Large Interactive Displays}

\author{Patrick Reipschl{\"a}ger*, Tamara Flemisch*, Raimund Dachselt
}
\authorfooter{

\vspace{-1em}
\begin{adjustwidth}{-12pt}{}
  * The first two authors contributed equally to this work.

  All authors are with the Interactive Media Lab at the Technische Universit{\"a}t Dresden, Germany.
  Raimund Dachselt is with the Centre for Tactile Internet (CeTi) and the Cluster of Excellence Physics of Life, TU Dresden.
  E-mails: \{patrick.reipschlaeger, tamara.flemisch\}@tu-dresden.de, dachselt@acm.org.
\end{adjustwidth}

}

\shortauthortitle{Reipschl{\"a}ger \MakeLowercase{\textit{et al.}}: Personal Augmented Reality for Information Visualization on Large Interactive Displays}

\input{content/abstract}

\keywords{Augmented Reality, Information Visualization, InfoVis, Large Displays, Immersive Analytics, Physical Navigation, Multiple Coordinated Views}

\CCScatlist{ 
 \CCScat{K.6.1}{Management of Computing and Information Systems}%
{Project and People Management}{Life Cycle};
 \CCScat{K.7.m}{The Computing Profession}{Miscellaneous}{Ethics}
}

\teaser{
  \centering
  \includegraphics[width=\linewidth]{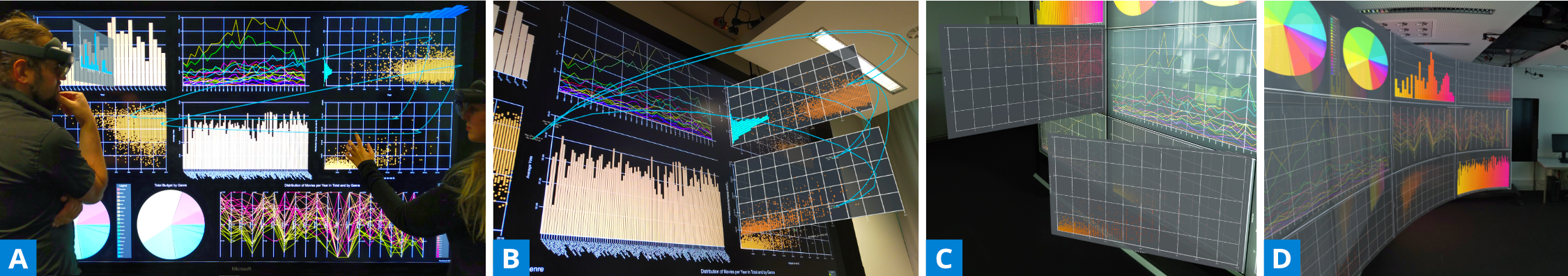}
  \vspace{-1.5em}
  \caption{Impressions from our prototype for extending visualizations with Augmented Reality: (a) Two analysts working with our environment, (b) display showing our \vtBrushLink, \vtEmbedded, and \vtExtendedViews, (c) \vtHinged to improve perception of remote content, and (d) \vtCurved providing an overview of the entire display.}
	\label{fig:teaser}
}

\vgtcinsertpkg

\begin{document}


\firstsection{Introduction}\label{sec:intro}

\maketitle


\input{content/introduction}
\input{content/related-work}

\input{content/generalConcept}

\input{content/techniques}
\input{content/prototype}
\input{content/useCase}

\input{content/discussion}
\input{content/conclusion}

\acknowledgments{
  We want to thank Mats Ole Ellenberg, Severin Engert, and Remke Dirk Albrecht for their support in creating this publication.
  This work was funded by
  DFG grant 389792660 as part of TRR~248 (see \url{https://perspicuous-computing.science}) and
  the DFG as part of Germany's Excellence Strategy Clusters of Excellence EXC-2068 390729961 \textit{Physics of Life}
  and EXC 2050/1 390696704 \textit{Centre for Tactile Internet with Human-in-the-Loop (CeTI)} of TU Dresden.
}

\newpage
\bibliographystyle{abbrv-doi-hyperref}
\bibliography{ms}
\end{document}

%% file: customization.tex
\usepackage{xcolor}
\usepackage{xspace}
\usepackage{enumitem}
\usepackage{blindtext}
\usepackage{soul}

\definecolor{NoteBlue}{RGB}{0,0,191}
\newcommand\note[1]{{\color{NoteBlue}{(Note: #1)}}}

\definecolor{TodoRed}{RGB}{225,63,63}
\newcommand\todo[1]{{\color{TodoRed}{(ToDo: #1)}}}

\definecolor{ControversialGreen}{RGB}{0,127,0}

\definecolor{UnfinishedBlue}{RGB}{0,127,127}

\definecolor{PriorGreen}{RGB}{42,127,80}
\definecolor{RevisionPurple}{RGB}{159,84,254}

\definecolor{ReferenceOrange}{RGB}{255,116,59}

\definecolor{preipPurple}{RGB}{118, 22, 248}

\definecolor{maraTeal}{RGB}{0, 180, 220}

\renewcommand{\note}[1]{}
\renewcommand{\todo}[1]{}

\sethlcolor{yellow}
\definecolor{revisionBlue}{RGB}{0, 138, 255}
\newcommand\cl[1]{{\color{revisionBlue}{#1}}}
\renewcommand{\cl}[1]{#1}

\newcommand*{\imana}{\emph{Immersive Analytics\nocorr}\xspace}

\newcommand{\iconChallengesPercep}{\icon{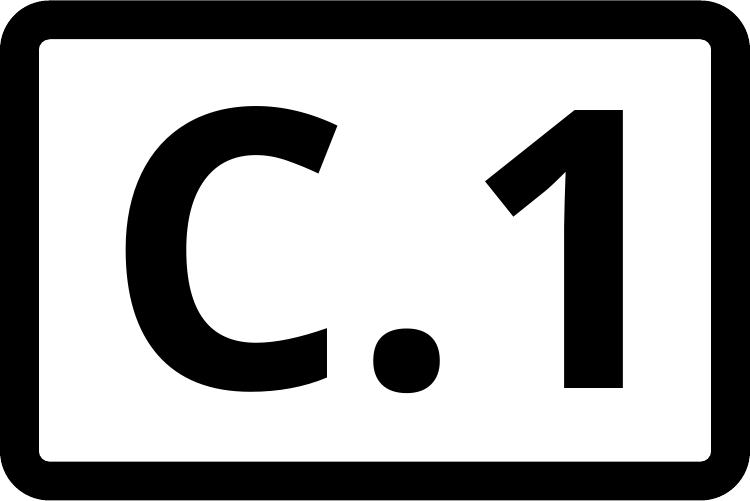}}

\newcommand{\iconChallengesComplex}{\icon{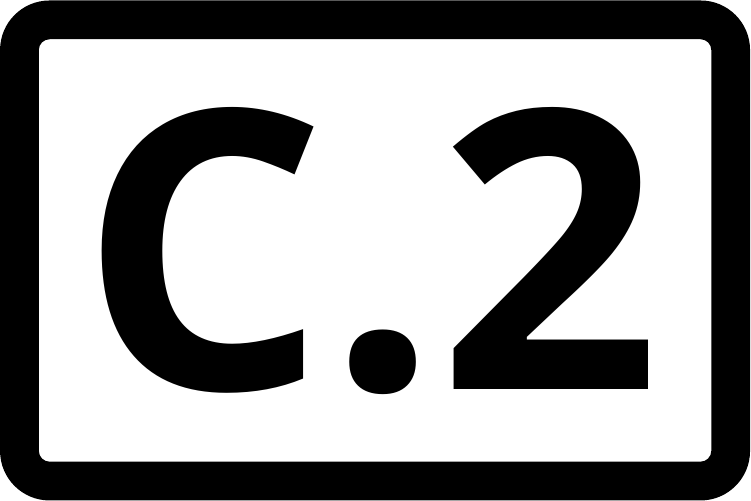}}

\newcommand{\iconChallengesMultiUser}{\icon{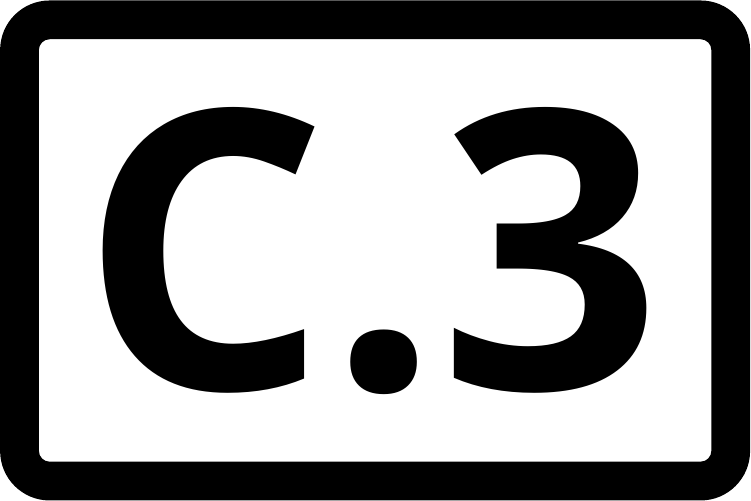}}

\newcommand{\pPercep}{\iconChallengesPercep}
\newcommand{\pPercepName}{Perceptual Issues on Large Displays}

\newcommand{\pComplex}{\iconChallengesComplex}
\newcommand{\pComplexName}{Managing Density and Complexity}
\newcommand{\pComplexFull}{\pComplex \pComplexName}

\newcommand{\pMultiUser}{\iconChallengesMultiUser}
\newcommand{\pMultiUserName}{Support for Multiple Users}

\newcommand{\vt}[1]{\emph{#1}\xspace}
\newcommand{\vtHinged}{\vt{Hinged Visualizations}}
\newcommand{\vtCurved}{\vt{Curved AR Screen}}
\newcommand{\vtEmbedded}{\vt{Embedded AR Visualizations}}

\newcommand{\vtExtendedViews}{\vt{Extended Axis Views}}
\newcommand{\vtMagicLenses}{\vt{Magic AR Lenses}}
\newcommand{\vtAnnotate}{\vt{AR Annotations}}
\newcommand{\vtBrushLink}{\vt{AR Brushing and Linking}}
\newcommand{\vtLayer}{\vt{AR Visualization Layers}}

\newcommand{\icon}[1]{\includegraphics[height=0.8em]{#1}\xspace}
\newcommand{\iconSpatialZonesCenter}{\icon{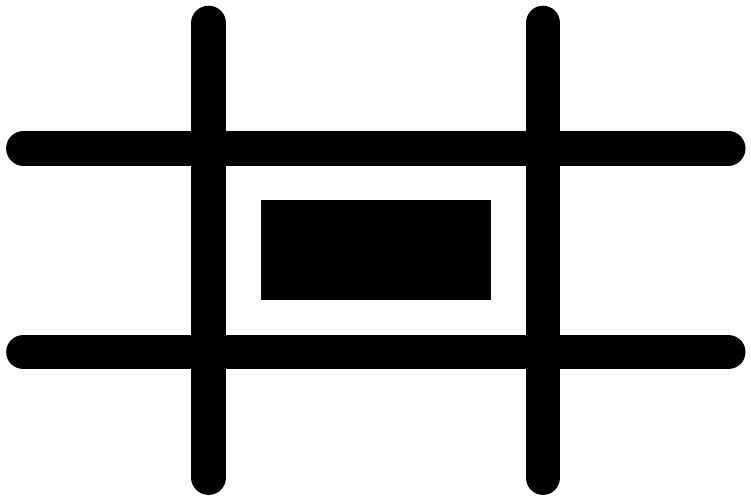}}
\newcommand{\iconSpatialZonesLeft}{\icon{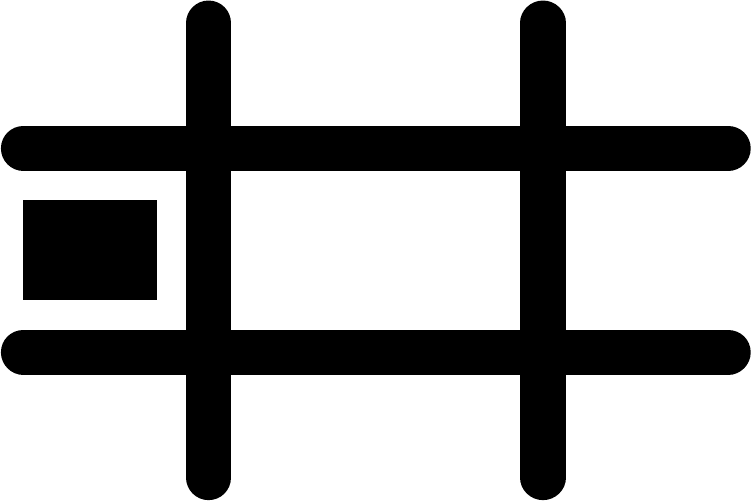}}
\newcommand{\iconSpatialZonesRight}{\icon{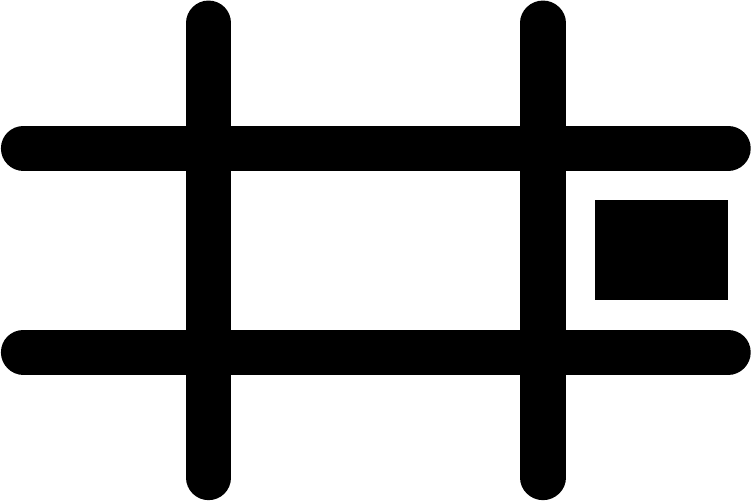}}
\newcommand{\iconSpatialZonesTop}{\icon{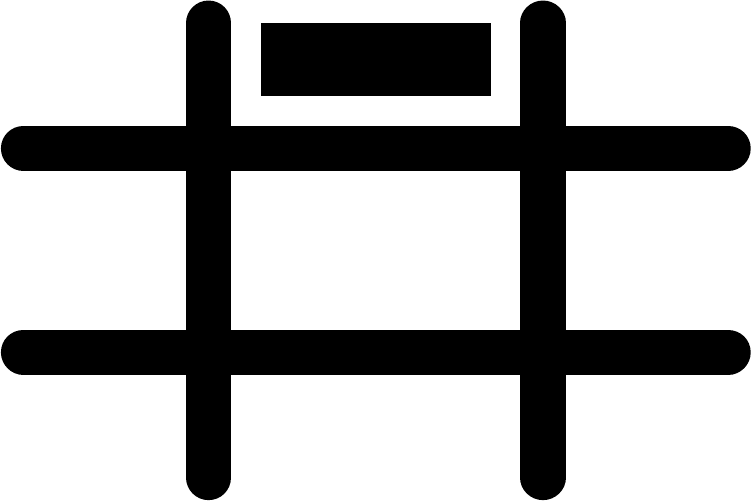}}
\newcommand{\iconSpatialZonesBottom}{\icon{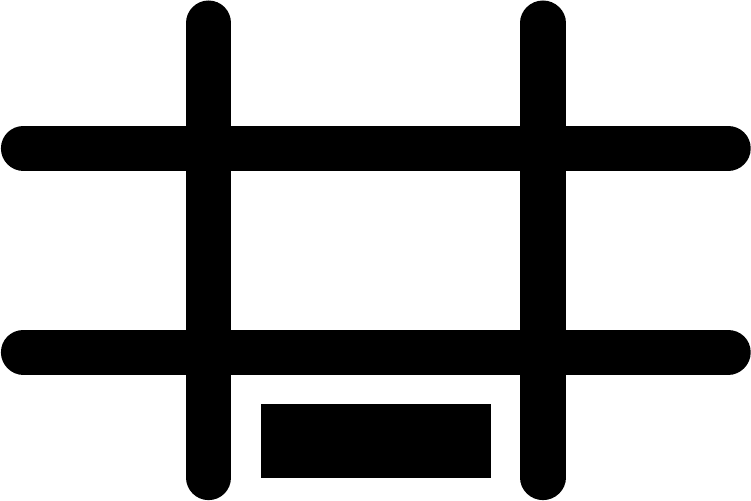}}
\newcommand{\iconSpatialZonesTopBottom}{\icon{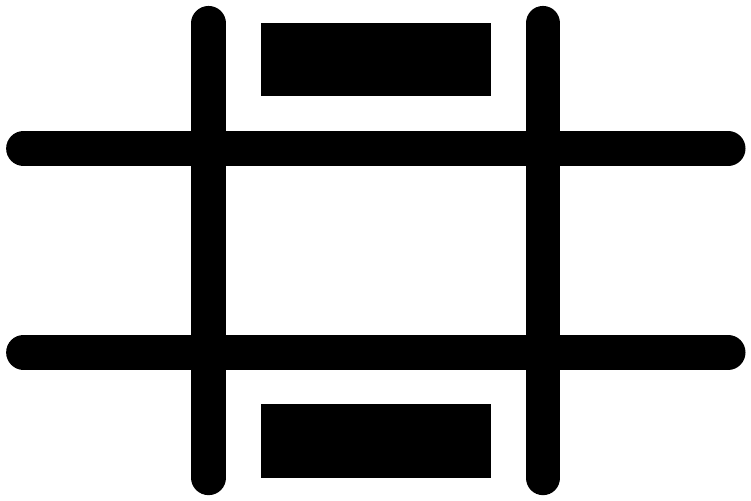}}
\newcommand{\iconSpatialZonesLeftRight}{\icon{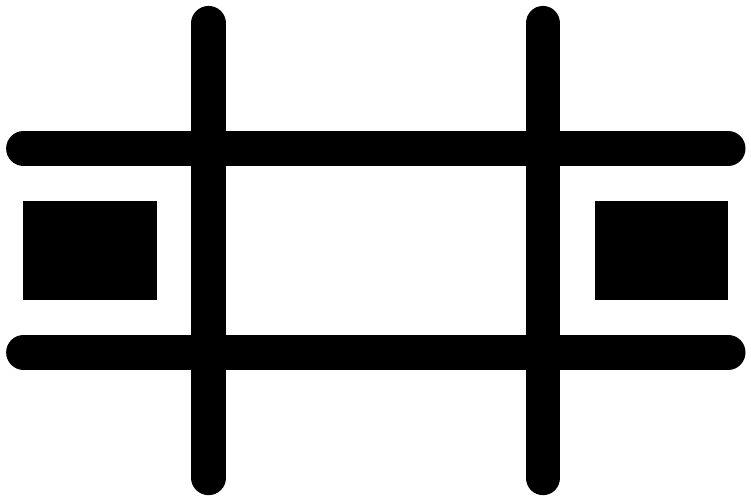}}
\newcommand{\iconSpatialZonesAll}{\icon{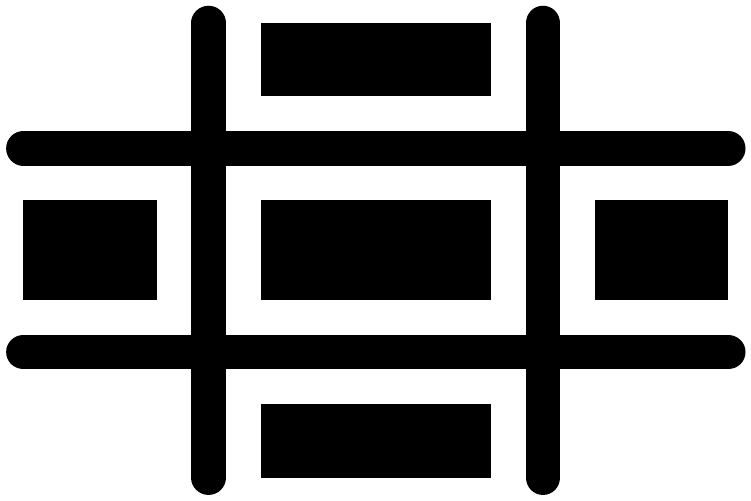}}

\newcommand{\src}{\iconSpatialZonesCenter}
\newcommand{\srl}{\iconSpatialZonesLeft}
\newcommand{\srr}{\iconSpatialZonesRight}
\newcommand{\srt}{\iconSpatialZonesTop}
\newcommand{\srb}{\iconSpatialZonesBottom}

\newcommand{\iconVisView}{\icon{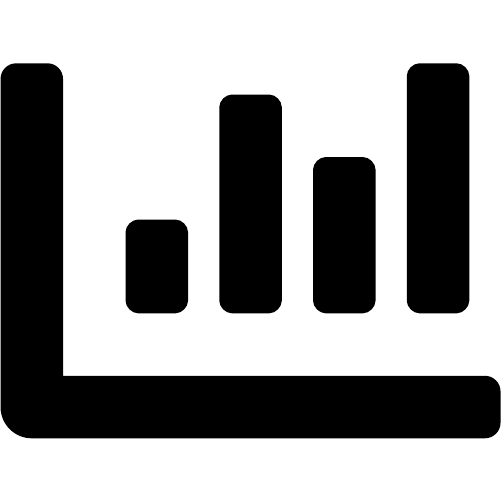}}
\newcommand{\iconVisTools}{\icon{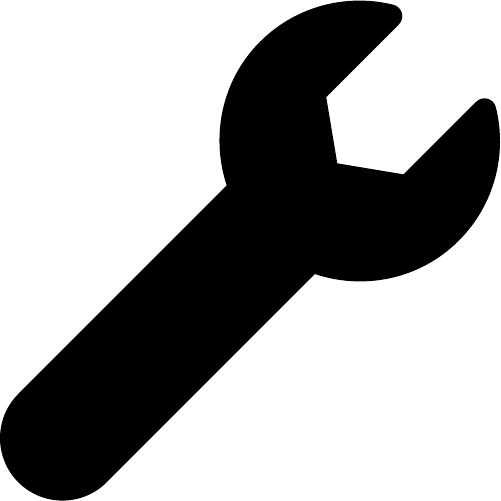}}
\newcommand{\iconVisSelections}{\icon{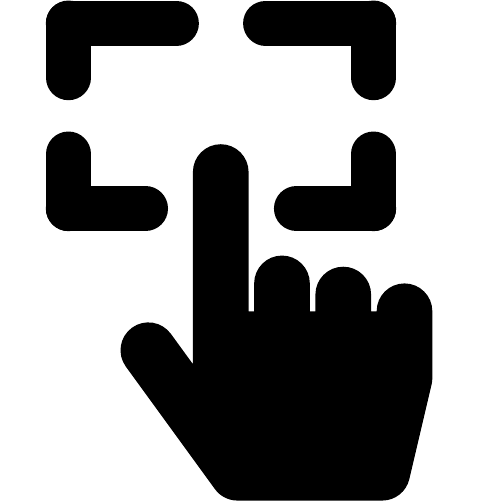}}
\newcommand{\iconVisAxes}{\icon{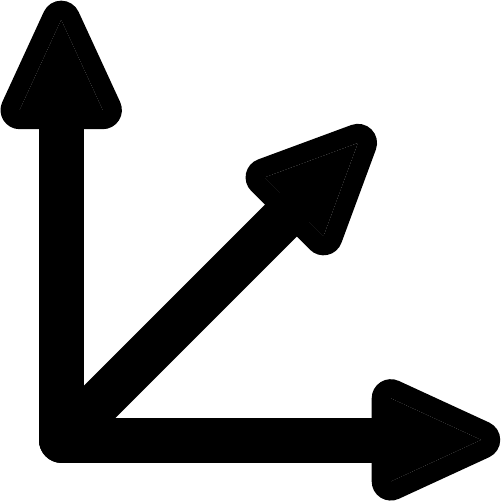}}
\newcommand{\iconVisLinks}{\icon{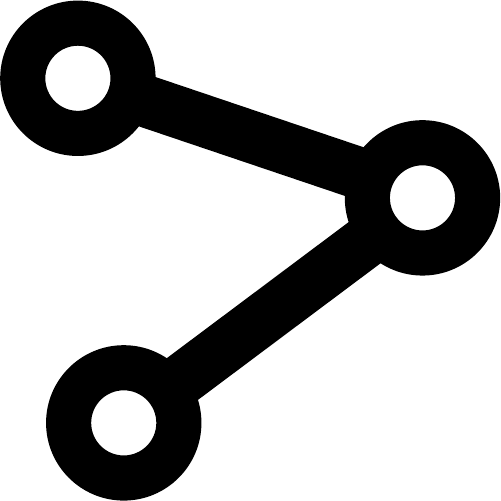}}
\newcommand{\iconVisMarks}{\icon{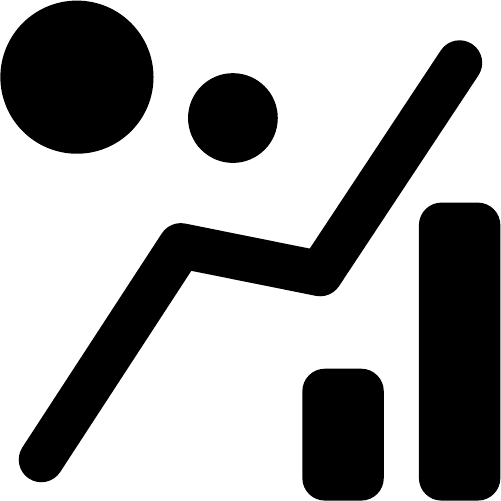}}
\newcommand{\iconVisAnnotations}{\icon{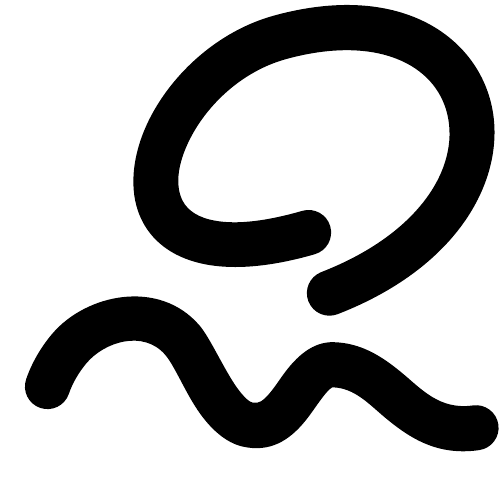}}
\newcommand{\iconVisLegend}{\icon{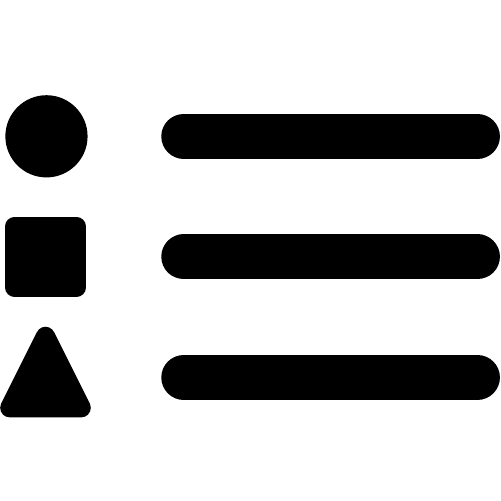}}
\newcommand{\iconVisUIElements}{\icon{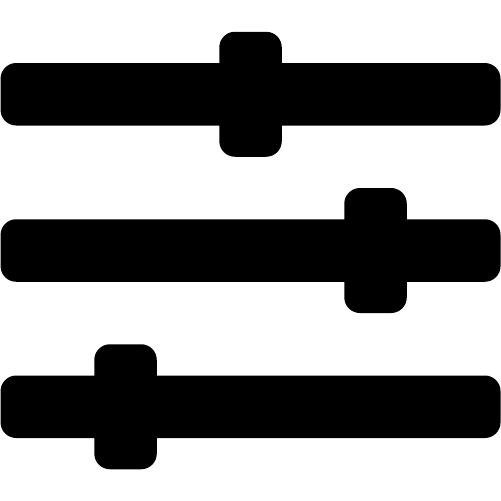}}

\newcommand{\vAxes}{\iconVisAxes}
\newcommand{\vMarks}{\iconVisMarks}
\newcommand{\vLinks}{\iconVisLinks}

\newcommand{\vLegend}{\iconVisLegend}
\newcommand{\vAnnotations}{\iconVisAnnotations}
\newcommand{\vSelections}{\iconVisSelections}
\newcommand{\vUIElements}{\iconVisUIElements}
\newcommand{\vTools}{\iconVisTools}
\newcommand{\vView}{\iconVisView}

\newcommand{\iconPersViewsOcclusion}{\icon{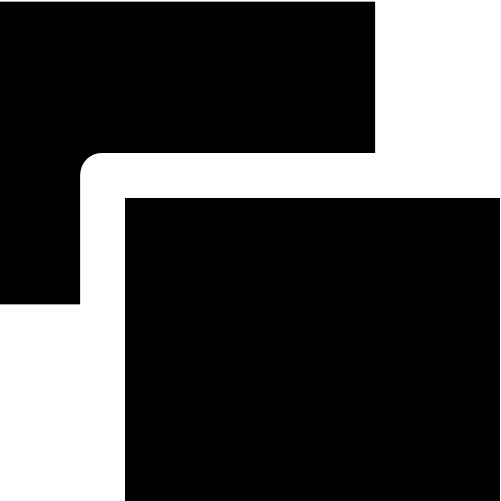}}
\newcommand{\iconPersViewsPersonalized}{\icon{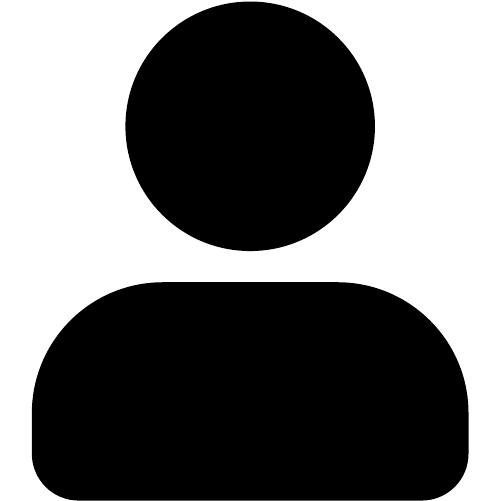}}
\newcommand{\iconPersViewsPrivate}{\icon{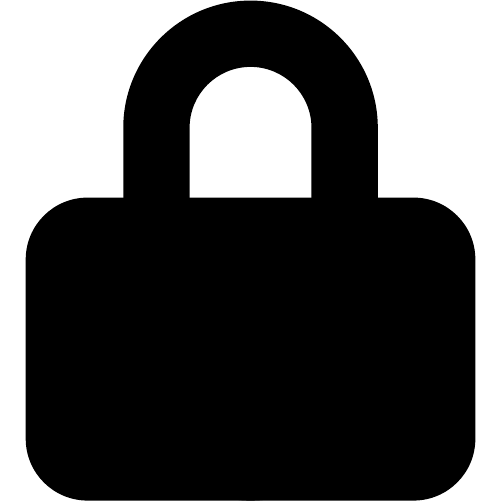}}

\newcommand{\userX}{Sue\xspace}
\newcommand{\userY}{Matt\xspace}

\newcommand*{\fullref}[1]{\hyperref[{#1}]{\autoref*{#1} \nameref*{#1}}}

\newcommand\refx[2]{(see~\autoref{#1}#2)\xspace}

\newcommand\emphref[1]{\emph{\nameref{#1}}\xspace}

\newcommand*{\parainline}[1]{{\vspace{0.5em}\noindent\fontsize{8pt}{8pt}\selectfont\sffamily\bfseries#1}\hspace{3pt}}

\definecolor{lorem}{RGB}{224,224,224}

%% file: content/abstract.tex
\abstract{In this work we propose the combination of large interactive displays with personal head-mounted Augmented Reality (AR) for information visualization to facilitate data exploration and analysis.
Even though large displays provide more display space, they are challenging with regard to perception, effective multi-user support, and managing data density and complexity.
To address these issues and illustrate our proposed setup, we contribute an extensive design space comprising first, the spatial alignment of display, visualizations, and objects in AR space.
\cl{Next, we discuss which parts of a visualization can be augmented.}
Finally, we analyze how AR can be used to display personal views in order to show additional information and to minimize the mutual disturbance of data analysts.

Based on this conceptual foundation, we present a number of exemplary techniques for extending visualizations with AR and discuss their relation to our design space.
We further describe how these techniques address typical visualization problems that we have identified during our literature research.
To \cl{examine} our concepts, we introduce a generic AR visualization framework as well as a prototype implementing several example techniques.
In order to demonstrate their potential, we further present a use case walkthrough in which we analyze a movie data set.
From these experiences, we conclude that the contributed techniques can be useful in exploring and understanding multivariate data.
We are convinced that the extension of large displays with AR for information visualization has a great potential for data analysis and sense-making.
} 

%% file: content/introduction.tex

Interactive information visualization on large or wall-sized displays has been shown to have great potential for data analysis~\cite{Andrews2010,Horak2018,Langner2019}.
Due to their size, they can visualize large amounts of data or several visualizations at once which are often linked to each other. 
Such multiple coordinated views are useful to help understanding complex, multivariate data sets by highlighting connections between views through Brushing and Linking~\cite{Roberts2007}.
However, working in front of large displays also leads to awareness and perception issues regarding the peripheral areas of the display~\cite{Andrews2011,Bezerianos2012}.
This \cl{results} in users having to step back from the display to gain an overview of the data~\cite{Andrews2011,Langner2019} which in turn means that details within specific visualizations are no longer perceivable.
While multiple coordinated visualizations are a valuable technique for analyzing multivariate data sets, they still require extensible context switches.
Therefore, it can be beneficial to visualize all available information directly in place.
Due to their size, large displays are also suited for collaborative work~\cite{Isenberg2012,Jakobsen2014} enabling several users to interact with the display at the same time.
However, as users interact with the display, they block parts of it with their body which \cl{can disturb other analysts}~\cite{Peltonen2008,Jakobsen2016}.
Highlights, which result from Brushing and Linking, can also be distracting to others.
The same is true for other people's tools that might obstruct important details of a visualization.

Besides large displays, the \cl{recent} field of \imana has shown great potential for improving data analysis and mitigating problems commonly associated with three dimensional visualizations on traditional displays~\cite{Whitlock2020}.
By using immersive technologies, \imana aims to overcome barriers between people, their data, and the tools they use for analysis and decision making~\cite{Book:ImAn2018}.
In this regard, physical navigation~\cite{Jansen2019,Ball2007,Endert2011}, spatial immersion~\cite{Andrews2010}, and embodied data exploration~\cite{Kister2015,Whitlock2020} have proven to help in understanding complex data~\cite{Book:ImAn2018}.
\imana applications also have limitations regarding interaction and the visual quality of visualizations~\cite{Bach2018a,Book:ImAn2018,Batch2020}.
By combining \imana and large displays, their individual limitations can be mitigated and their advantages strengthened~\cite{Bach2018a}.

Based on these promising two avenues of research, we propose to combine a large, interactive display with head-mounted, personal Augmented Reality (AR).
We imagine one or more analysts in front of a large, vertical, interactive display which shows one large or several linked visualizations of multivariate data.
Each analyst wears an AR head-mounted display (HMD) which is used to overlay the screen with additional, user-specific information.
We are convinced that this combination can mitigate the aforementioned issues for both, large displays and pure immersive analysis techniques.
The display serves as a frame of reference for all AR content and facilitates natural pen and touch interaction\cite{Walny2012}.
Augmented Reality, however, can help to guide analysts' awareness, show remote parts of the screen, and to display additional content directly embedded in the visualization~\cite{Willett2017}.
Furthermore, this enables personal views which foster undisturbed and unobstructed interaction and allow for individual, personalized selections, annotations, and extensions to the data shown on the display. 
In this work, we contribute an analysis of the design space of the combination of a large, interactive display, Augmented Reality, and information visualizations.
In this design space, we first analyze the spatial alignment of the display, the analyst, and the AR content.
\cl{Second, we discuss which parts of a visualization can be augmented.}
Third, we describe how to leverage personal views to show additional information and minimize the mutual disturbance of analysts.
To illustrate the potential of our conceptual design space, we contribute several augmentation techniques, which address specific visualization problems.
\cl{To examine our concepts}, we developed a generic, data-driven AR visualization framework as well as a prototype implementing our proposed techniques \refx{fig:teaser}{}.
We demonstrate their potential in a use case scenario in which we analyze a movie data set.
Lastly, we discuss technical and perceptual aspects and give an outlook on how the presented concepts can be extended in the future.

%% file: content/related-work.tex
\section{Related Work and Background}\label{sec:rw}

Our work touches upon the research fields of Information Visualization (InfoVis) and Human-Computer Interaction (HCI).
In particular, it refers to the use of (1) large, interactive displays and (2) Mixed Reality (MR) for information visualization as well as (3) the general combination of interactive surfaces and Augmented Reality. 

\subsection{Information Visualization on Large Displays}

Large, interactive displays have been increasingly used for data analysis and visualization with positive impact~\cite{Andrews2010,Rajabiyazdi2015}.
Their screen size, the vast physical space in front of them~\cite{Ball2007,Jakobsen2015,Andrews2010}, and their interaction capabilities make them a powerful foundation for exploring large amounts of data.
Possible interaction modalities for information visualization beyond the traditional desktop setup have been proposed~\cite{Lee2012,Roberts2014}, with touch interaction being the most common one.
Data analysis benefits from touch input due to a more direct interaction with the data~\cite{Chegini2019,Horak2018,Langner2019,Kister2017,Jakobsen2014,Prouzeau2017}, although some use cases profit from more distant interaction~\cite{Kister2015,Kister2017,Langner2019}.
The increased screen estate of the displays can be used for showing large amounts of data, such as node-link diagrams~\cite{Prouzeau2017,Kister2015}, or to incorporate multiple views of data~\cite{Langner2019,Lee2019}.
Due to their size, large displays also facilitate collaborative data analysis and multiple analysts working in parallel~\cite{Isenberg2012,Tobiasz2009,Prouzeau2017,Kister2015,Langner2019}.
Even though the benefits of large, interactive surfaces for visualization is evident, they also contribute specific challenges, affordances, and requirements~\cite{Andrews2011} which we further discuss in~\autoref{sec:challenges}.

\subsection{Immersive Analytics}

The field of \imana makes use of various display technologies to facilitate and support embodied data analysis and decision making~\cite{Book:ImAn2018}.
Recent research in \imana focuses especially on how Mixed Reality environments can help to explore and understand data.
Cordeil et al.~\cite{Cordeil2017} present a system to explore multidimensional data in \cl{Virtual Reality (VR)} through manipulating and arranging linked axes.
Approaches on how to ideally route links in immersive spaces have been analyzed by Prouzeau et al.~\cite{Prouzeau:2019:LinkRouting}.
Alternatively, Liu et al.~\cite{Liu2020} adapted the concept of small multiples to an immersive environment.
Overall, different MR technologies have their own strengths and weaknesses~\cite{Bach2018a}.
In this regard, Augmented Reality shows great potential for data analysis since it allows for embedding data representations into the physical space and therefore facilitates the simultaneous exploration of large amounts of data~\cite{Willett2017,Batch2020}.

Whitlock et al.~\cite{Whitlock2020} studied how visualization techniques and different immersive display modalities affect data analysis tasks.
They found that analysts benefit from AR and VR when estimating depth and height but also that color perception in video see-through AR is challenging.
The impact of different edge types for node-link diagrams in AR was investigated by B\"{u}schel et al.~\cite{Bueschel:2019:LinkStyles}.
Batch et al.~\cite{Batch2020} found that analysts are highly interested, present, and engaged when analyzing data in immersive environments. 
Similar to visualization on large displays, \imana also presents new affordances and requirements regarding the perception, design, and interaction of and with visualizations~\cite{Book:ImAn2018,Whitlock2020}.

Interaction in immersive environments is often achieved through mid-air gestures.
Although they allow for rich input modalities, they lack haptic feedback~\cite{Book:ImAn2018}.
Therefore, recent work has combined AR with multi-touch displays.
Wang et al.~\cite{Wang2020} propose to integrate Augmented Reality visualizations into traditional desktop setups.
Their findings suggest that this helps analysts to better understand their data.
Work by Chen et al.~\cite{Chen2020} and \emph{VisAR} by Kim et al.~\cite{Kim2017} both combine personal AR views with small as well as large displays to generate interactivity for otherwise static visualizations.
Sun et al.~\cite{Sun2019} incorporate visualizations on wall-sized displays and AR HMDs for collaborative visual analysis and personalized views.
Likewise, Butcher et al.~\cite{Butscher2018} apply a video see-through AR HMD to augment a large interactive tabletop for analyzing 2D visualizations.
AR visualizations are directly situated on the display and analysts interact with proxies on the interactive surface.
Furthermore, work that shows display-like surfaces in a pure VR environment~\cite{Filho2019,Filho2020,Prouzeau:2019:LinkRouting} may be adapted to a setup consisting of a physical display and head-mounted AR.

\subsection{Combining Interactive Surfaces with Mixed Reality}

Starting with the early works \emph{Hybrid Displays}~\cite{Feiner1991} and \emph{Augmented Surfaces}~\cite{Rekimoto1999}, the recent development of commercial VR and AR HMDs has led to a renewed interest in combining Mixed Reality with interactive or non-interactive displays.
Bodgan et al.\cite{Bogdan:2014} and Mendes et al.~\cite{Mendes:2014} use a stereoscopic display to investigate the differences between 2D and 3D interaction for modeling and 3D manipulation of objects, but there is no consensus which technique is preferable.
Grubert et al.~\cite{Grubert:2015} use head-mounted Augmented Reality to extend the screen around mobile devices.
\emph{DualCAD} by Milette and McGuffin~\cite{Millette2016} loosely combines a traditional desktop CAD application with head-mounted AR.
In \emph{SymbiosisSketch} by Arora et al.~\cite{Arora2018} users can draw objects in AR using a tablet.
However, the interaction on the tablet is spatially decoupled from the AR surface that is sketched onto.
In a similar way, Sereno et al.~\cite{Sereno:2019} propose to manipulate an AR 3D visualization by using touch on a tablet.
Again, there is no spatial connection between the tablet and the visualization.
\cl{Cavallo et al.'s \emph{DataSpace}~\cite{Cavallo2019} combines multiple large displays with AR and VR visualizations but with no strong spatial integration between content in AR and on the displays (also cf. \cite{Cavallo2019a}).}
Reipschl{\"a}ger and Dachselt\cite{Reipschlaeger2019} combine an interactive surface with an AR HMD for a stereoscopic 3D modeling tool.
In this work, there is a strong connection between the display and the AR content and a particular focus is on how the borders of the display can be utilized to place additional views and how to interact with them.
They have also coined the term \emph{Augmented Displays} for seamlessly combining interactive displays with head-mounted AR, to which our approach can also be attributed.

\subsection{Challenges for Visualization on Large Displays}\label{sec:challenges}

Based on previous works on visualization for large, interactive displays, we compiled several important challenges, which we describe in detail in the following sections.

\subsubsection*{{\Large \pPercep} \pPercepName}\label{sec:challenges:percep}
Large displays have special requirements regarding human perception due to their size.
Endert et al.'s~\cite{Endert2011} list of perceptual issues with large displays consists of the importance of peripheral vision, distortion due to extreme viewing angles and long distances, and the different effects of visual variables on large displays compared to traditional displays.
\cl{\textbf{Change Blindness:}}
Change blindness is a big problem on large displays due to the screen extending into peripheral vision~\cite{Endert2011} which makes acquiring an overview of the data~\cite{Perteneder2016} or comparisons \cite{Andrews2011} much more difficult.
Motion and animation can be used to create awareness in peripheral areas~\cite{Bartram2003}, e.g., by temporary flashing, afterglow effects~\cite{Baudisch2006}, or ambient light around the display~\cite{Perteneder2016}.
However, while they increase awareness, they do not improve perception of the data itself and users have to move to the actual position to see what exactly changed.
\cl{\textbf{Distortion due to Viewing Angles:}}
Extreme viewing angles can distort the shape~\cite{Wigdor2007} and orientation~\cite{Endert2011} of data marks as well as the perception of aspect ratios and color encodings~\cite{Andrews2011,Bezerianos2012}.
In contrast, others found color to be more robust to distortion~\cite{Yost2007,Endert2011} as well as length~\cite{Bezerianos2012} and position~\cite{Wigdor2007}.
Dynamically correcting the user's perspective~\cite{Nacenta2007}, images with multiple scales~\cite{Isenberg2013a}, lenses~\cite{Kister2015,Kister2017}, and potentially using curved displays~\cite{Prouzeau2016a} have been explored to address the issue of distortion.
\cl{\textbf{Physical Navigation:}}
Users move naturally in front of large displays~\cite{Ball2007,Rajabiyazdi2015} to utilize the increased space~\cite{Andrews2010}.
This physical navigation~\cite{Ball2007} usually seems to improve performance compared to virtual navigation~\cite{Jansen2019} and increases spatial memory~\cite{Andrews2010,Jakobsen2015}.
Generally, users step back from the display to gain an overview of the data~\cite{Ball2007,Endert2011} and move closer to access details~\cite{Bezerianos2012,Endert2011,Isenberg2013a}.
\cl{\textbf{Effects of Visual Encodings:}}
However, perceptual issues emerge when stepping back from the display as visual encodings will be lost at a distance or are perceived differently~\cite{Yost2007,Endert2011,Isenberg2013a}.
This can be mitigated by visual aggregation through using color or creating light and dark regions~\cite{Endert2011,Yost2007}.
Proxemics~\cite{Ballendat2010}, i.e., adapting the information displayed according to a user's proximity, is another technique to handle this issue, e.g., by adjusting the level of detail~\cite{Ledo2015,Jakobsen2013} or by using geometrical zooming~\cite{Kister2015}.
Hybrid-image visualization~\cite{Isenberg2013a} takes another direction by embedding different levels of detail into one visualization which are only visible from a particular distance.
Another strategy is offloading detail views on additional devices that can be used from afar~\cite{Kister2017,Horak2018}.

\subsubsection*{{\Large \pComplex} \pComplexName}\label{sec:challenges:complex}
The size of large displays is their most important characteristic but at the same time also represents the challenge how the obtained space is best used.
Andrews et al.~\cite{Andrews2011} provide several approaches to using the available space to increase the amount of displayed information.
\cl{\textbf{Appropriate Use of Space:}}
While just increasing the number of data items is an obvious option~\cite{Sun2019,Chegini2019}, adding more data dimensions~\cite{Tobiasz2009} might result in a more comprehensive visualization.
Further details, like labels and values~\cite{Agarwal2019,Sun2019} and multiple levels of data~\cite{Langner2015}, can be embedded or incorporated into the visualization.
Beside scaling visualizations and using the space to enhance one visualization, the acquired space can be used to show multiple, coordinated visualizations~\cite{Roberts2007,Langner2019,Agarwal2019,Chegini2019,Tobiasz2009}.
However, given a big enough data set, even large screens might not suffice to display them\cite{Elmqvist2008a,Ion2013}.
Therefore, it is important to consider alternative ways of visualizing additional data besides increasing the screen space.
\cl{\textbf{Scaling of Visual Encodings:}}
Managing the density and complexity of data is not a problem exclusive to visualization on smaller screens, but poses an issue on large displays as well.
Some visual encodings, such as the number of data marks and their size~\cite{Andrews2010,Yost2007} as well as color itself~\cite{Endert2011}, scale well with size and benefit from the additional space. Others, such as the number of perceivable colors and network connectivity~\cite{Andrews2011}, i.e., edge crossings, do not improve with additional space.
\cl{\textbf{Overplotting:}}
Problems such as overplotting also remain on large displays.
For scatter plots, clutter reduction, displacement, zooming, and density plots~\cite{Sarikaya2018} as well as pixel-based techniques~\cite{Raidou2019} can be used to mitigate them.
For dense node-link diagrams, edge bundling~\cite{Lhuillier2017} and cutting~\cite{Edge2018} can be used to minimize edge crossings and reduce edge congestion.
\cl{\textbf{Data Complexity:}}
Other common approaches to reduce the complexity of data are clustering and dimensionality reduction~\cite{Liu2017, Wenskovitch2018} as well as utilizing focus + context techniques, e.g., magic lenses~\cite{Tominski:2017:MagicLenses} or space-folding~\cite{Elmqvist2008a}, overview + detail techniques~\cite{Ion2013}, and semantic zooming~\cite{Kister2017}.
For zoomed information spaces, off-screen visualization techniques~\cite{Baudisch2003,Gustafson2008,Petford2019} and offloading information to auxiliary devices to increase screen estate~\cite{Kister2017,Horak2018} are also beneficial.

\subsubsection*{{\Large \pMultiUser} \pMultiUserName}\label{sec:challenges:multiuser}
Large displays are naturally suitable for multi-user scenarios.
Their large size and the thereby defined physical space on the screen and in front of it makes them ideal for co-located work and collaboration~\cite{Jakobsen2014,Kister2015,Isenberg2011}.
\cl{\textbf{Obstruction and Distraction:}}
Large displays also lead to challenges as users will inevitably physically obstruct, disturb, and distract each other when working in groups~\cite{Jakobsen2014}.
Prouzeau~et~al.~\cite{Prouzeau2017} suggest to leverage interactions that have a large visual footprint to encourage close collaboration and facilitate the frequent switch~\cite{Jakobsen2014} from closely to loosely coupled collaboration~\cite{Tang2006}.
Furthermore, users can easily lose awareness for what others are doing while working on a different area of the large display~\cite{Gutwin2002,Tang2006,Jakobsen2014,Mateescu2019}.
Gutwin \& Greenberg~\cite{Gutwin2002} state that to maintain workspace awareness, users need to monitor each other's behavior, for instance by showing past interactions of others~\cite{Hajizadeh2013}, incorporating visual connections between information~\cite{Isenberg2012}, using ambient light~\cite{Perteneder2016}, or embedding animation and off-screen visualization techniques~\cite{Petford2019,Gustafson2008}.
\cl{\textbf{Personal Territories:}}
Studies have shown that users value a personal territory to interact with while still maintaining shared spaces~\cite{Tang2006,Scott2004}.
On vertical displays, territories are mostly transient and dynamic~\cite{Jakobsen2014} when moving around and nonetheless similar to those on horizontal displays~\cite{Azad2012}.
Having a personal space can either be achieved by a designated space on the display itself or an additional personal device, such as a smartphone~\cite{Kister2017,Chegini2019} or a smart watch~\cite{Horak2018}.

\vspace{1em}\noindent
So far, we discussed issues for visualizing data on large, interactive displays and common strategies for addressing them.
These presented strategies might sufficiently resolve a particular challenge.
However, we now introduce a novel approach and a promising alternative to address many of these challenges and potentially mitigate shortcomings of already existing techniques.

%% file: content/generalConcept.tex
\section{General Concept and Design Space}\label{sec:concept}
Our goal with this work is to address the aforementioned challenges and improve the visualization of multivariate data on large, interactive displays.
We assume a scenario consisting of one or multiple data analysts in front of a large, vertical, interactive display that either shows a single large visualization or multiple coordinated views (MCV) of multivariate data.
We propose to equip each of these analysts with an AR HMD that extends visualizations or parts of visualizations with personalized information.
As a delimitation to other work, we solely focus on setups consisting of a single large display and exclude multi-display environments (cf. \cite{Kister2017,Ledo2015}).
We only consider vertically oriented screens of larger size, e.g., digital whiteboards up to wall-sized displays.
Consequently, we assume that the display remains stationary during use and does not change its position.
\cl{Our concepts can be used in a multi-user environment and we discuss cases with both single and multiple users.
However, we do not focus on such techniques fostering communication and collaboration between users.}

\input{content/designSpace}

%% file: content/designSpace.tex

\cl{We contribute a design space that provides a systematic overview defining and exploring the boundaries of our proposed setup.
It can be used to inspire and guide the development of new visualization and interaction techniques.}
\cl{It consists of (1) the spatial alignment of the display, the users, and the AR content, (2) how different parts of a visualization can be augmented, and (3) how to provide personal content in multi-user scenarios.}

\subsection{Spatial Alignment of Display and AR Content}\label{sec:ds:spatialrelation}
The placement of AR content with respect to its spatial alignment to the display is an essential design aspect for augmenting visualizations.
In this section, we present possible placement options and outline their respective implications.
Reipschl{\"a}ger and Dachselt~\cite{Reipschlaeger2019} have previously defined three levels of spatial proximity that relate to our design space: AR objects situated directly on or in front of the display, objects close or directly at the edge of the display, and objects with no spatial relation to the display.
Our work focuses mostly on the first two levels, as we consider a strong connection and a precise spatial alignment of AR and display content to be of essential importance.

\begin{figure*}[htbp]
	\begin{minipage}[t]{0.24\textwidth}
		\includegraphics[width=\textwidth]{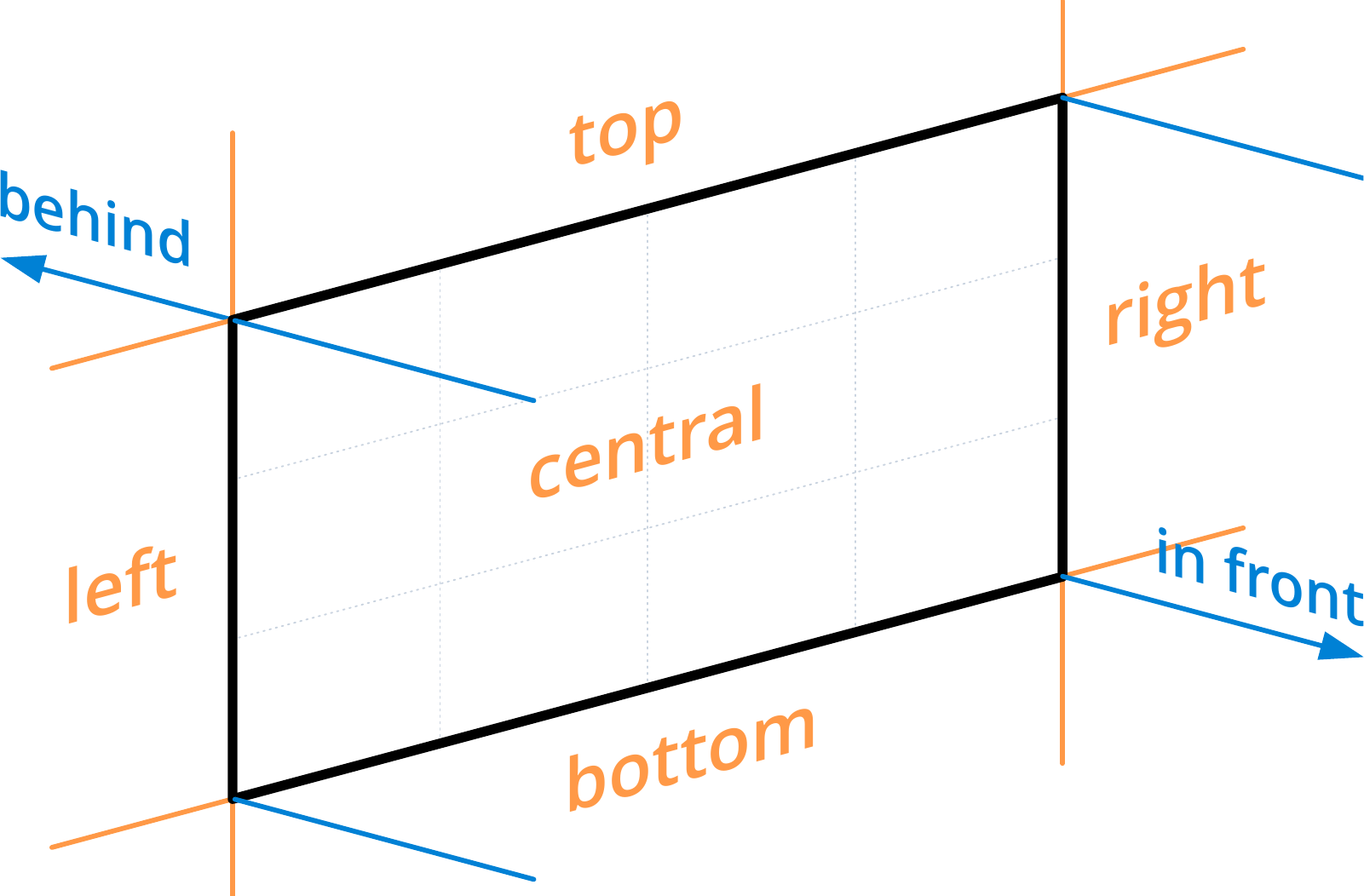}
		\caption{Planar zones of spatial alignment of AR content with the display, which extend in front and behind the display's screen.}
		\label{fig:ds:spatialRelation}
	\end{minipage}%
	\hspace{0.01\textwidth}
	\begin{minipage}[t]{0.37\textwidth}
		\includegraphics[width=\textwidth]{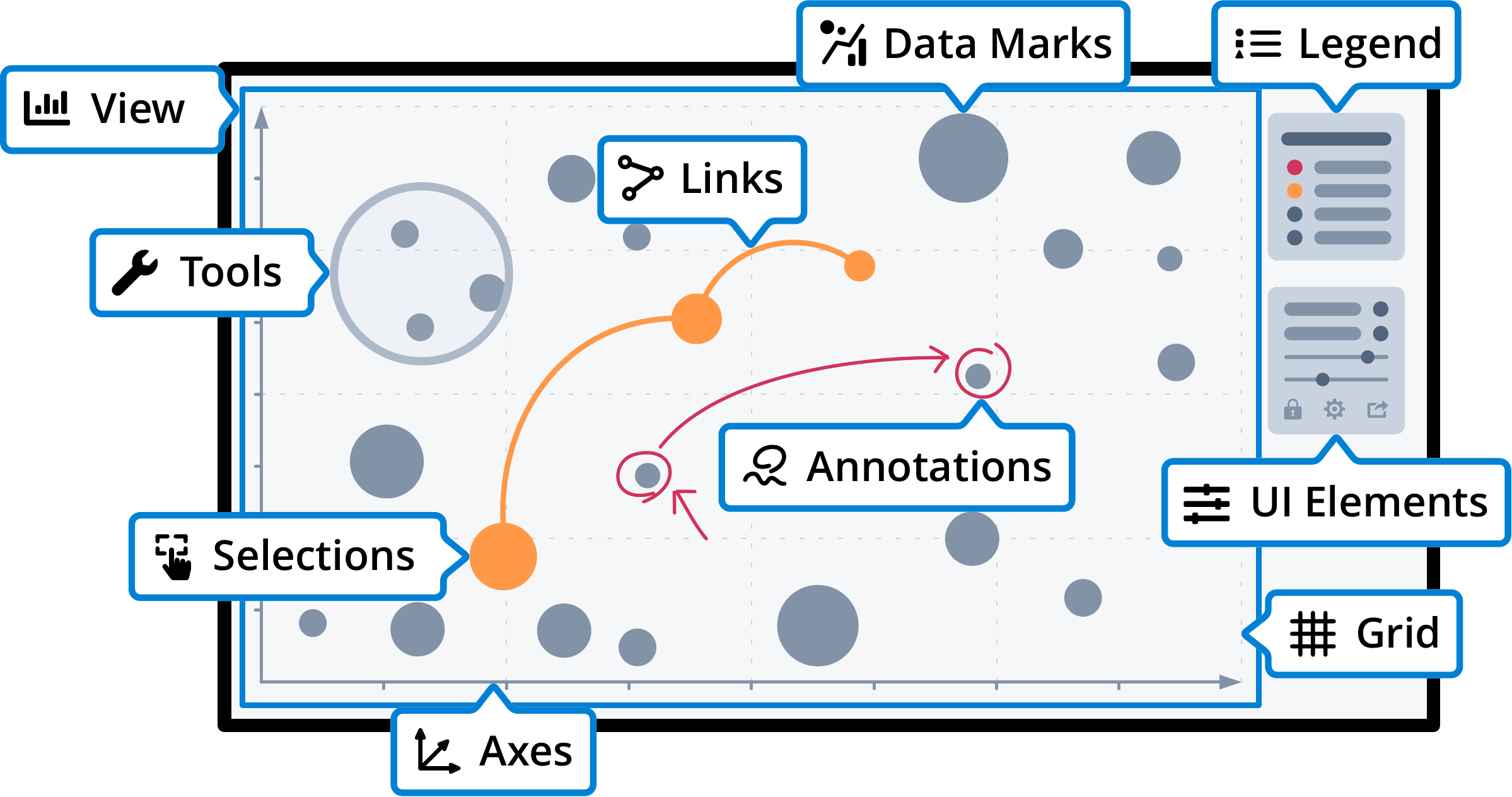}
		\caption{Overview of the components of a visualization that can be extended with AR content \cl{and the symbols used throughout the paper to represent those components}.}
		\label{fig:ds:components}
	\end{minipage}%
	\hspace{0.01\textwidth}
	\begin{minipage}[t]{0.36\textwidth}
		\includegraphics[width=\columnwidth]{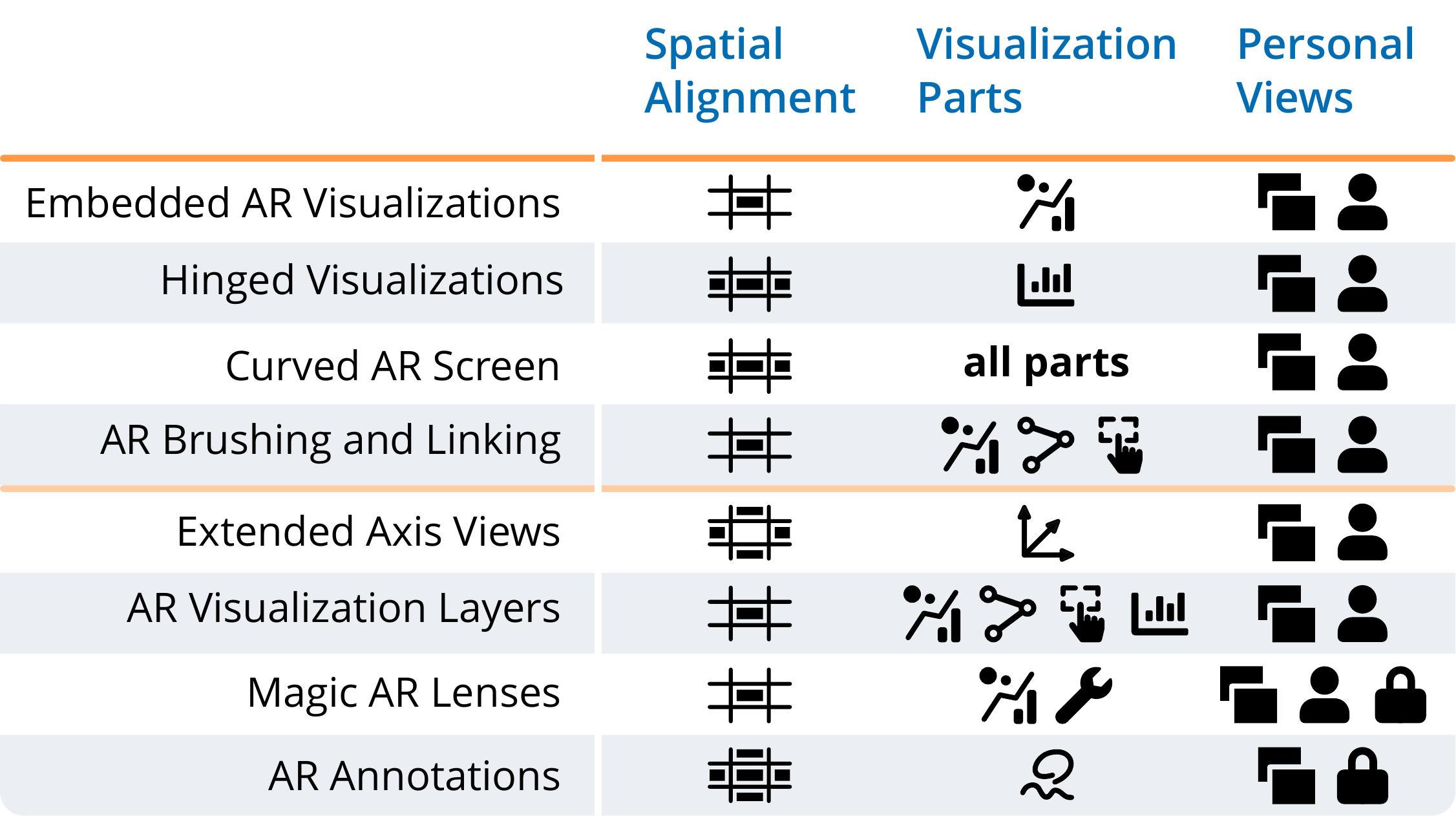}
		\caption{\cl{Overview of our design space dimensions and their relation to the techniques presented in \autoref{sec:techniques}}.}
		\label{fig:ds:summary}
	\end{minipage}%
	\vspace{-1em}
\end{figure*}

We classify the display's screen as a frame of reference for the placement of AR objects, defining a planar space around the screen.
Content positioned on this plane represents a natural extension of the screen and thus conveys a strong relation to the display.
It can be divided based on the cardinal directions to create nine different principal zones to place objects, which can be further extended orthogonally into the third dimension \cl{for placing content in front or behind the display}\footnote{
While these spatial zones are considered primarily for the placement of AR content in the context of this work, they also describe the general relationship of visualizations to their surrounding environment, regardless of the medium used.} \refx{fig:ds:spatialRelation}{}.
Although in principle, each of the zones is suitable for placing arbitrary AR content, there are special relationships that should be considered.
Of all zones, the \cl{central zone~\src} is the most important one because it has the strongest alignment to the display and to content on the display.
For visualizations with a rectangular shape, the \cl{left zone~\srl} and \cl{right zone~\srr} are particularly suitable for content that relates to the vertical dimensions of the visualization.
Consequently, the \cl{top zone~\srt} and \cl{bottom zone~\srb} are particularly suitable for content that refers to the horizontal dimensions.
An example for this is to extend a data dimension shown in the visualization on the screen to the left and right or top and bottom to provide additional data.
Since the four corner zones have the weakest spatial connection to the display, the other zones should be preferred for placing objects.
\cl{AR content may also occupy more than one zone, e.g., top and bottom~\iconSpatialZonesTopBottom, left and right~\iconSpatialZonesLeftRight, or even all of them~\iconSpatialZonesAll.
Furthermore, Content on, behind, or in front of a spatial zone is considered to be within that specific zone.}
With increasing display size, all spatial zones besides the central zone become less important because of their increasing distance to the center of the screen and therefore also to the likely position of the analysts.
\cl{Conversely, the smaller the display gets, the more important the peripheral zones become.
We are therefore convinced that these spatial zones are also useful for the design of techniques for smaller screens.}
The current position of an analyst in relation to the display can also influence the spatial positioning of the AR content (cf. Proxemic Interaction~\cite{Ballendat2010}).
For example, if an analyst is closer to the left edge of the display than to the right, AR content can be displayed preferably in \cl{the left zone compared to the right zone.}
In practice, some zones may be unsuitable for placement of AR content if they conflict with or are constrained by the physical condition of the room.
To avoid this, the spatial zones can be adapted to the room geometry, e.g., using the floor of the room for the \cl{bottom zone}.

\subsection{Augmentations for Parts of a Visualization}\label{sec:ds:partsOfVis}

While the previous considerations could be applied to every type of content, this section concentrates on how visualizations themselves can be augmented.
A visualization consists of components, such as \cl{axes~\vAxes}, \cl{data marks~\vMarks}, and labels, that are common to almost every visualization and which can be extended using AR techniques (see \autoref{fig:ds:components} for a comprehensive list).
In general, augmenting specific parts of a visualization can address problems specific to large displays, which we describe in \autoref{sec:challenges}.
Especially data marks benefit vastly from using the space around and in front of them to present additional, potentially multivariate data.
We can harness the property that every analyst has their own personal view to accommodate for \cl{legends~\vLegend}, labels, tooltips, personal tools as well as \cl{selections~\vSelections}.
\cl{Annotations~\vAnnotations} of the data can be created by, e.g., using a digital pen to write on the display.
Another common issue in visualization is overplotting and occlusion~(\pComplex) through data marks, for instance in dense scatter plots, or nodes and \cl{links~\vLinks} in node-link diagrams.
Using the third dimension, i.e., central zone \src, and incorporating Augmented Reality can help to address this issue by providing the increased space of another spatial dimension.
Additional \cl{UI elements~\vUIElements} for interaction with a visualization are fundamental for the functionality of interactive visualizations.
Therefore, we consider them as part of a visualization which can be augmented as well.

We have previously described how content in AR can be spatially aligned with the display.
However, this spatial alignment does not solely apply to a display but to individual visualizations and their components as well.
As a consequence, depending on the component, zones can either be very small (e.g., in case of a narrow \cl{axis~\vAxes}), restricted by other components in proximity (e.g., a bar that is next to another bar), or may not be available at all (e.g, a line covers only one dimension).
Larger zones are more suitable to be extended with AR content.
For instance, whereas the zones at the beginning and at the end of an axis can be used for auxiliary components, like UI elements, such as filters, labels, drop down menus, or legends, the adjacent and central zones of the axis are beneficial for displaying additional information for the visualization.
Furthermore, general UI elements or \cl{tools~\vTools} can be completely offloaded into AR if necessary and beneficial.
Besides augmenting parts of the visualization, augmentations can be used to enhance the \cl{visualization as a whole~\vView}, e.g., by highlighting recently updated visualizations or by extending the whole visualization into the third dimension.
Large displays are especially well suited for showing more than one visualization at a time due to their size~\cite{Andrews2011,Langner2019}.
As a consequence, the space around a single visualization may be constrained by other visualizations on the display (cf. \autoref{sec:ds:spatialrelation}).
To address this, the spatial zones can be folded by 90 degrees so that they are now orthogonal to the visualization.
Due to this spatial adaptation, the resulting augmentation space becomes similar to a cuboid.

\begin{figure*}[htbp]
	\begin{minipage}[t]{0.24\textwidth}
		\includegraphics[width=0.9\textwidth]{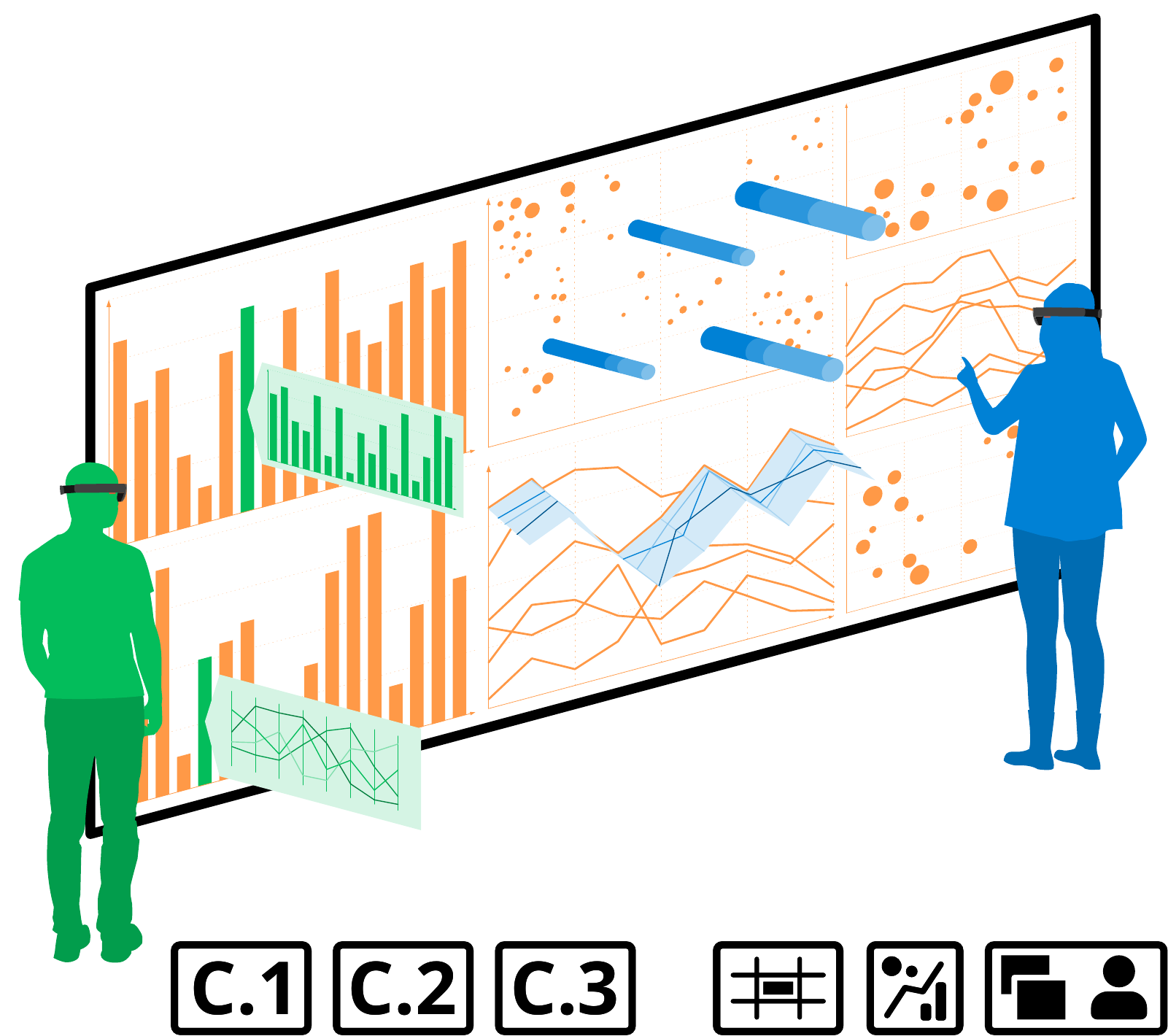}
		\caption{\vtEmbedded directly situated on a visualization on the display.}
		\label{fig:vt:embedded}
	\end{minipage}%
	\hspace{0.01\textwidth}
	\begin{minipage}[t]{0.24\textwidth}
		\includegraphics[width=0.9\textwidth]{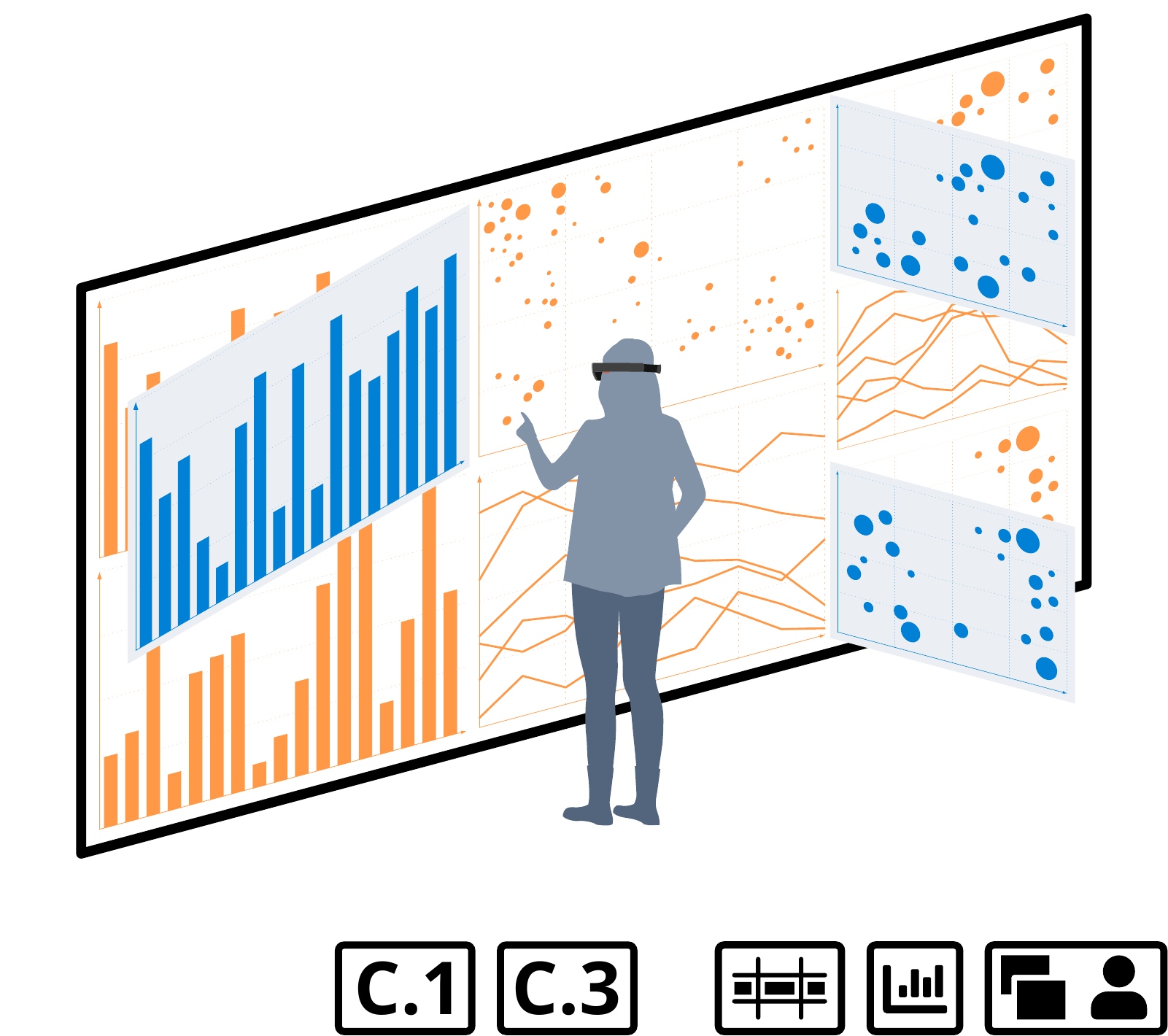}
		\caption{\vtHinged in AR space to address limited perception of far away objects.}
		\label{fig:vt:hinged}
	\end{minipage}%
	\hspace{0.01\textwidth}
	\begin{minipage}[t]{0.24\textwidth}
		\includegraphics[width=0.9\textwidth]{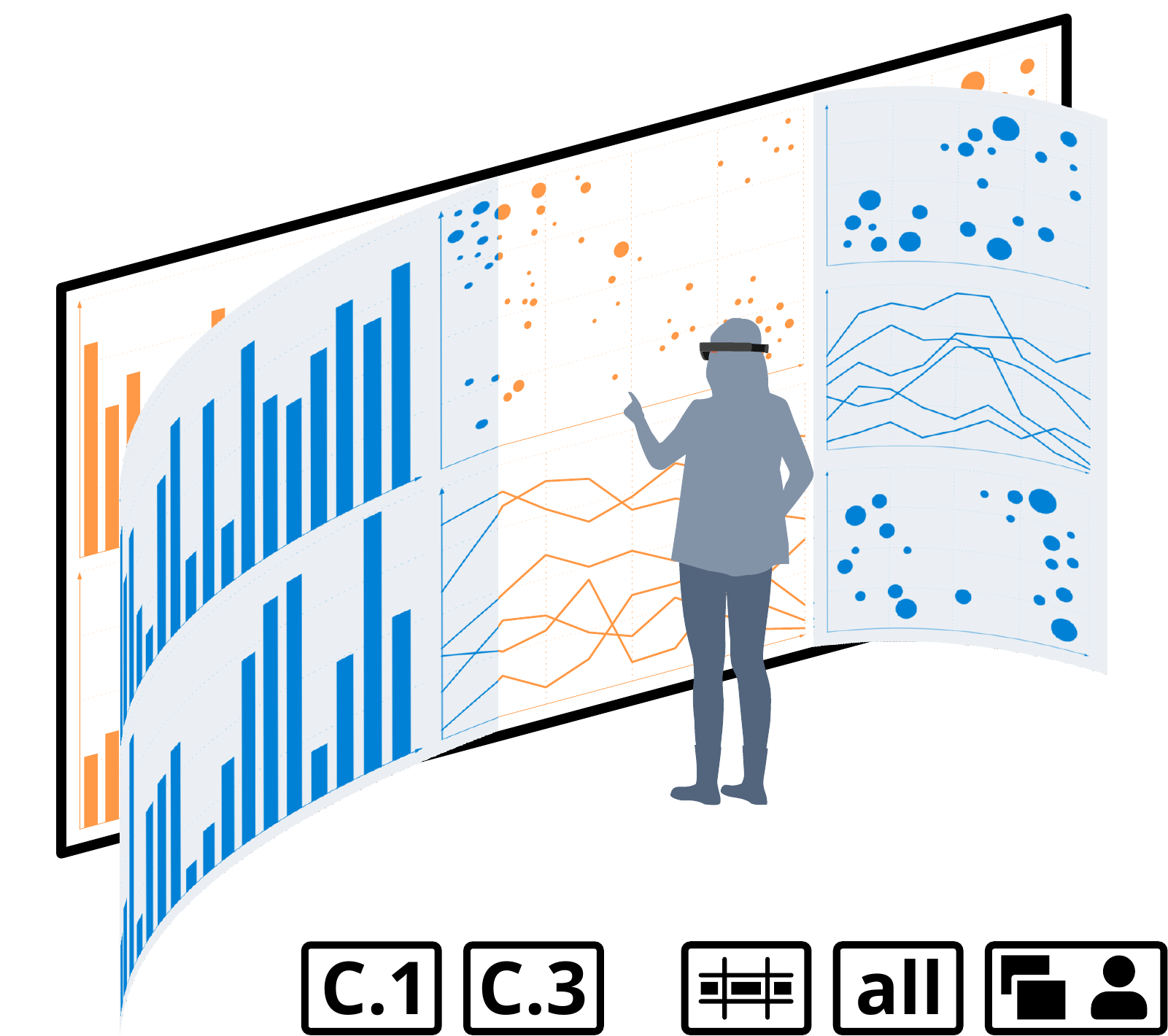} 
		\caption{\vtCurved to enable an overview of the whole display even when close to the screen.}
		\label{fig:vt:curved}
	\end{minipage}%
	\hspace{0.01\textwidth}
	\begin{minipage}[t]{0.24\textwidth}
		\includegraphics[width=0.9\textwidth]{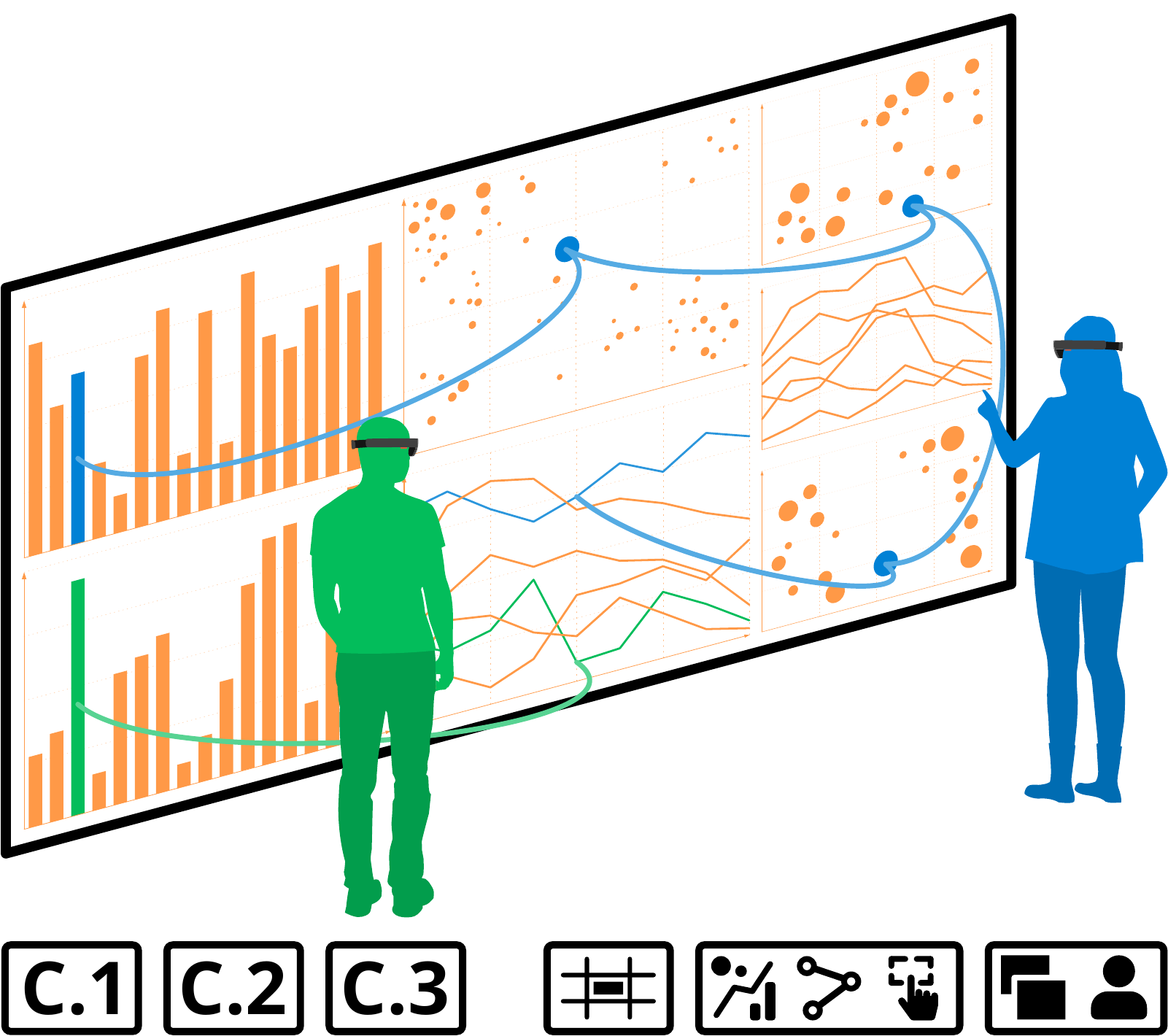}
		\caption{\vtBrushLink with 3D Beziér links indicating connections between visualizations.}
		\label{fig:vt:brushLink}
	\end{minipage}%
	\vspace*{-1.0em}
\end{figure*}

\subsection{Personal Views in Multi-User Scenarios}\label{sec:ds:personalViews}

\cl{
In addition to showing multiple visualizations, large displays are suitable for being used by multiple people at the same time~\cite{Andrews2011}.
This can result in several issues discussed in \pMultiUser~(\autoref{sec:challenges:multiuser}).
Combining displays with AR HMDs allows for \textbf{public views} as well as \textbf{personal views}.
By default, all content on the display can be seen by all users and all content on the AR HMDs is only perceptible by individual users.
However, this is more of a technological consequence and there is a distinction between content on the AR glasses being private because others simply cannot see it or content being private by design.
While it is difficult to hide content on the display for individual users, content in AR can be shared among multiple users to foster collaboration.
However, there can be benefits in not sharing AR content with others.
\textbf{\iconPersViewsOcclusion Preventing Obstruction:}
The additional views created by individual users might obstruct or confuse others.
An example is selections created by analysts which highlight data items in other visualizations through Brushing and Linking.
Since these highlights would be visible to everyone regardless of whether they are interesting for others, they should not be shared by default.
\textbf{\iconPersViewsPersonalized Personalized Views:}
Personalized views and tools can be configured to an individual analyst's needs.
For example, Magic Lenses~\cite{Tominski:2017:MagicLenses} are personal tools which can apply a variety of filters to a data set to support the current analyst's task.
Since these views are specific to every person, they should not be shared with others as they have their own personalized views.
\textbf{\iconPersViewsPrivate Private Information:}
People might consider certain information private and therefore want to keep it for themselves.
One example is annotations created by analysts which may contain content that people might feel uncomfortable in immediately sharing with others.
These aspects illustrate how AR can be used to enable personal views and in which cases they can be useful.
Of course, it is also important for collaborative work, how personal information and views can be shared with other analysts and how they can support social protocols and foster discourse.
We discuss this aspect in \autoref{sec:dis:collaboration}.
}

\vspace{0.8em}\noindent
\cl{The symbols we have introduced in this section for the different dimensions of the design space will be used in the conceptual illustrations for each technique to reference their relation to the design space. Additionally, \autoref{fig:ds:summary} provides an overview of the techniques and their associated parts in our design space.}

%% file: content/techniques.tex
\section{Visualization Techniques}\label{sec:techniques}

In this section, we present techniques that serve as examples to illustrate, explore, and give an overview of our design space.
We categorized our techniques by the challenges \pPercep and \pComplex depending on which of them is mainly addressed by a technique \refx{sec:challenges}{}.
Each technique also addresses challenge \pMultiUser as \cl{user-specific content is enabled through the use of personal AR HMDs}.
Due to space constraints, we opted to show the breadth of our proposed concepts by briefly describing a multitude of techniques instead of an in-depth analysis of a few techniques.
As a consequence, specific visualizations used to illustrate techniques are also exemplary and the same or similar augmentations can be potentially used for other suitable visualization techniques.
Furthermore, we will describe the associated interactions in general, instead of presenting a fine-grained interaction design.
Due to the combination of interactive displays and head-mounted AR, we envision the primary use of touch and pen interaction, both for the data and for controlling the individual techniques themselves.

\subsection{\pPercepName}
In this section, four techniques are presented that primarily address perceptual problems on large displays (\pPercep) and make it easier for analysts to better understand connections between the data and maintain an overview of the entire data set.

\subsubsection{Embedded AR Visualizations}\label{sec:vtEmbedded}
In contrast to a single visualization, multiple coordinated views are good at representing multivariate data sets. 
However, studies have shown that this is associated with many context and focus switches and that embedded visualizations may be preferable~\cite{Rajabiyazdi2015,Willett2017}.
We therefore propose to display additional data in AR space located directly on corresponding data objects of an existing visualization (see \autoref{fig:vt:embedded} and \autoref{fig:useCase_collage}c).
These \vtEmbedded extend orthogonally from the display.
In principle, any type of visualization can be embedded.
To ensure a strong spatial connection they can be adapted to the original form of the data marks they extend.
For example, a scatter plot can be extended by a cylinder-shaped visualization which encodes additional data dimensions similar to a stacked bar chart.
Due to the additional information displayed by the embedded visualization, reading the original visualization may become more difficult.
Therefore, it would be sensible to limit \vtEmbedded to data marks selected by analysts.
They can either be toggled individually with a touch tap or be shown on-demand for the duration they are touched.
\vtEmbedded address all three challenges, however primarily \pPercep by avoiding focus switches.
They also support \pComplex through their ability to incorporate multiple data levels and additional data into a single data marks.
\cl{Since they are user-specific, they can be configured by each analyst individually regarding what additional data is represented and how it is visualized (\pMultiUser).}
They use the \cl{central spatial zone} in front of a visualization.
We recommend to utilize the potential of the third dimension added by AR instead of simply showing another 2D visualization on a data mark.
As this technique is focused mainly on the extension of data marks, size needs to be considered.
The smaller a data mark is and the denser they are grouped together, the less space is available for showing embedded visualizations.

\subsubsection{Hinged Visualizations}\label{sec:vtHinged}
Perception issues through distortion caused by extreme viewing angles are a big problem on large displays (see \pPercep).
Therefore, we propose to support awareness for remote parts of the display by rotating these parts in AR space towards the analyst similar to a hinged door (see Fig.~\ref{fig:teaser}c and \ref{fig:vt:hinged}).
The larger the distance of an analyst towards the \vtHinged, the more it is tilted until it is aligned orthogonally to the display and faces the analyst.
Accordingly, as analysts move closer, the angle gradually decreases until the AR visualization has vanished to give way to the original visualization on the display.
Angles also decrease and the \vtHinged ultimately disappear the further \cl{the} analyst steps back from the display to perceive the visualizations as a whole.
\vtHinged are automatically generated based on the current selection of an analyst to allow for an overview and a rough comparison of connected values between remote \cl{visualizations}.
However, analysts can also explicitly toggle \vtHinged for specific visualizations regardless of their current selection.
A potential disadvantage of \vtHinged is that they can occlude other parts of the screen.
However, we consider this a minor issue since the occluded parts are difficult to detect by the current position of an analyst anyway.
\cl{Each analyst has individual \vtHinged based on their current data selection and position (\pMultiUser).}
\cl{\vtHinged take up the left, center, and right spatial zones and refer to the visualization as a whole.}
They are aimed at use cases where there is a large number of connected visualizations on the display.
It is only of limited assistance when few visualizations are present, since large parts of the screen have to be tilted.

\begin{figure*}[t]
	\begin{minipage}[t]{0.24\textwidth}
		\includegraphics[width=0.9\textwidth]{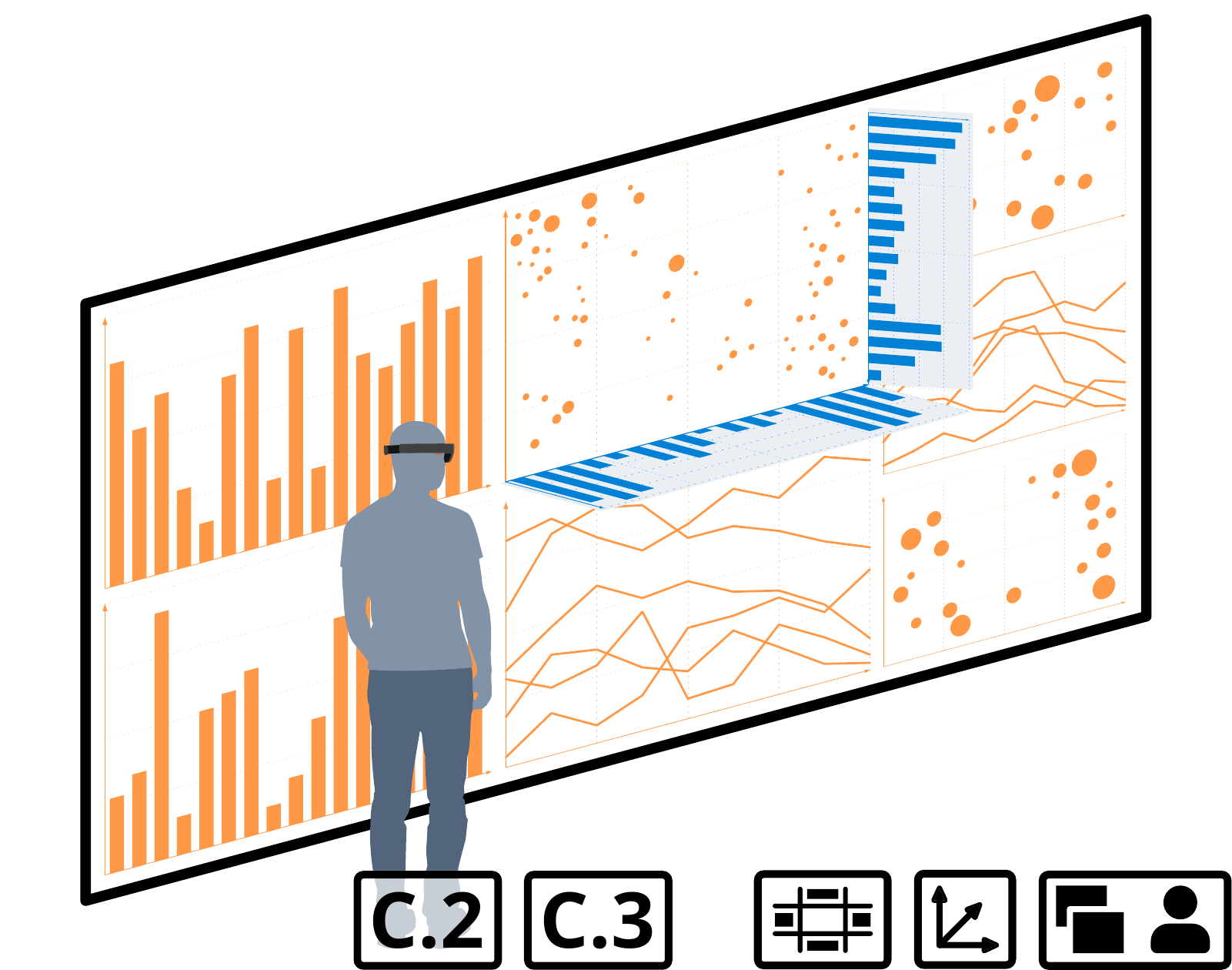}
		\caption{Expanding the axis of visualizations to create \vtExtendedViews and to show additional information, such as aggregations.}
		\label{fig:vt:extended}
	\end{minipage}%
	\hspace{0.01\textwidth}
	\begin{minipage}[t]{0.24\textwidth}
		\includegraphics[width=0.9\textwidth]{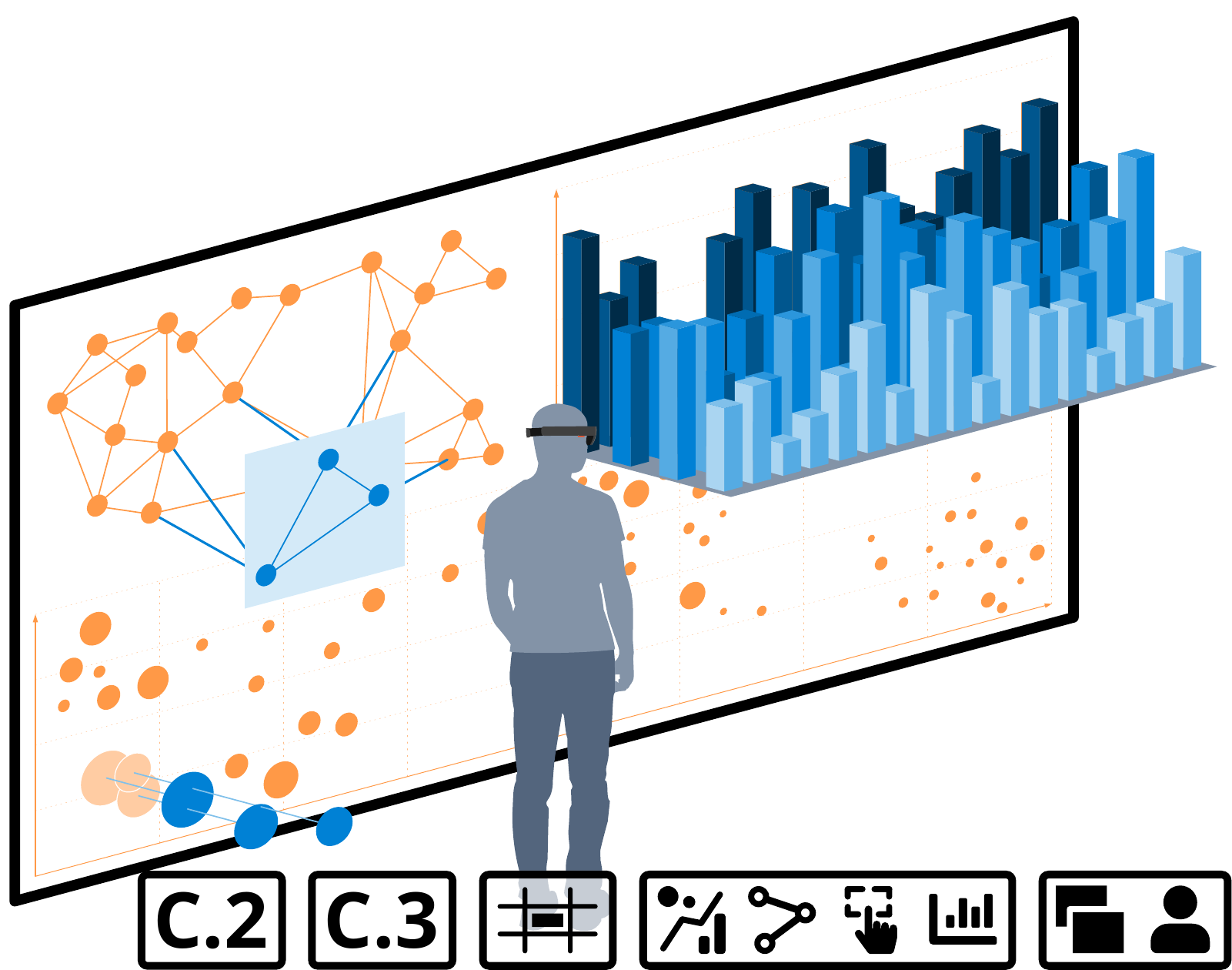}
		\caption{Showing \vtLayer in front of the display to offload data or to show additional data for an existing visualization.}
		\label{fig:vt:layers}
	\end{minipage}%
	\hspace{0.01\textwidth}
	\begin{minipage}[t]{0.24\textwidth}
		\includegraphics[width=0.9\textwidth]{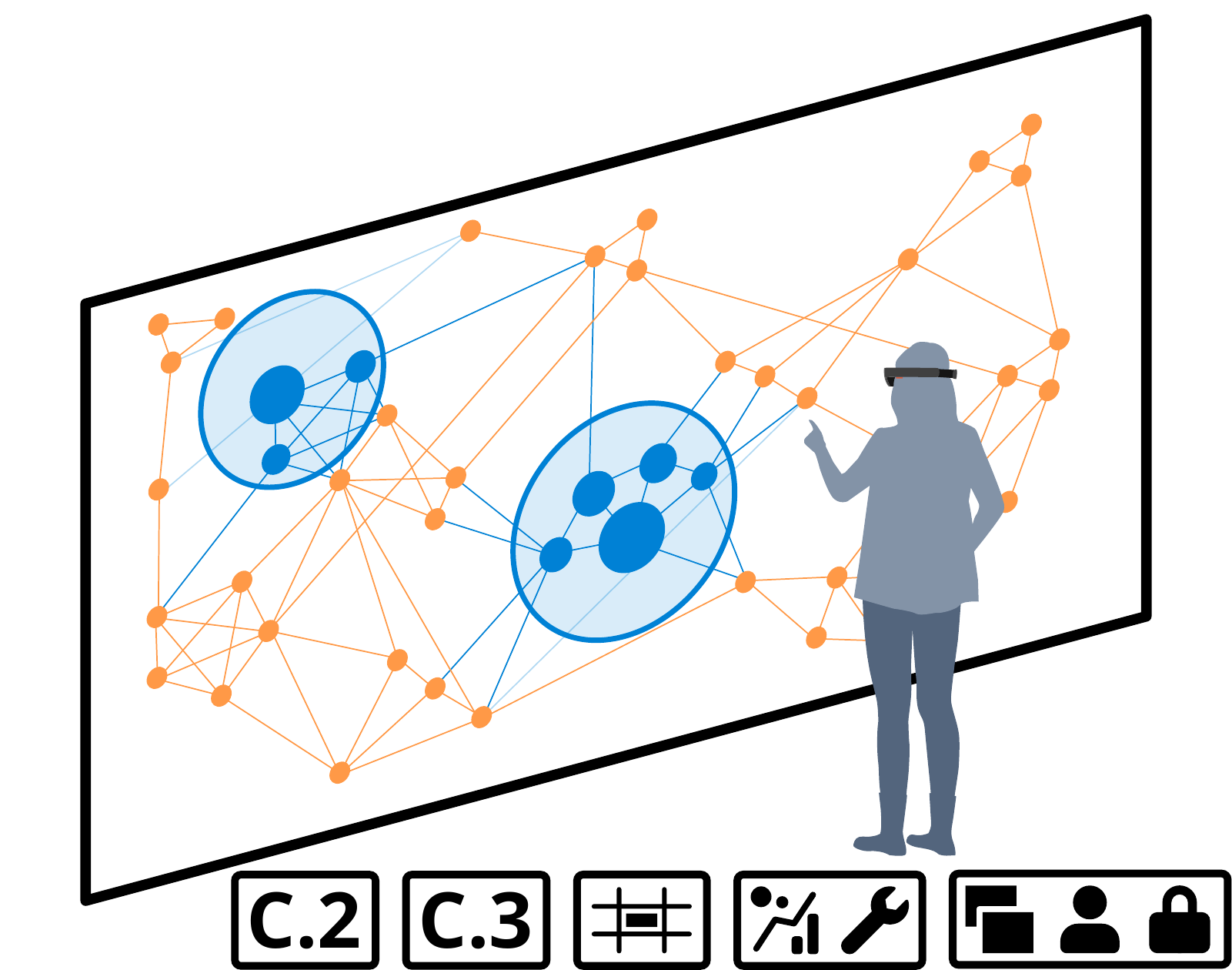}
		\caption{\vtMagicLenses as an example for personalized tools transferred into AR space.}
		\label{fig:vt:magicLenses}
	\end{minipage}%
	\hspace{0.01\textwidth}
	\begin{minipage}[t]{0.24\textwidth}
		\includegraphics[width=0.9\textwidth]{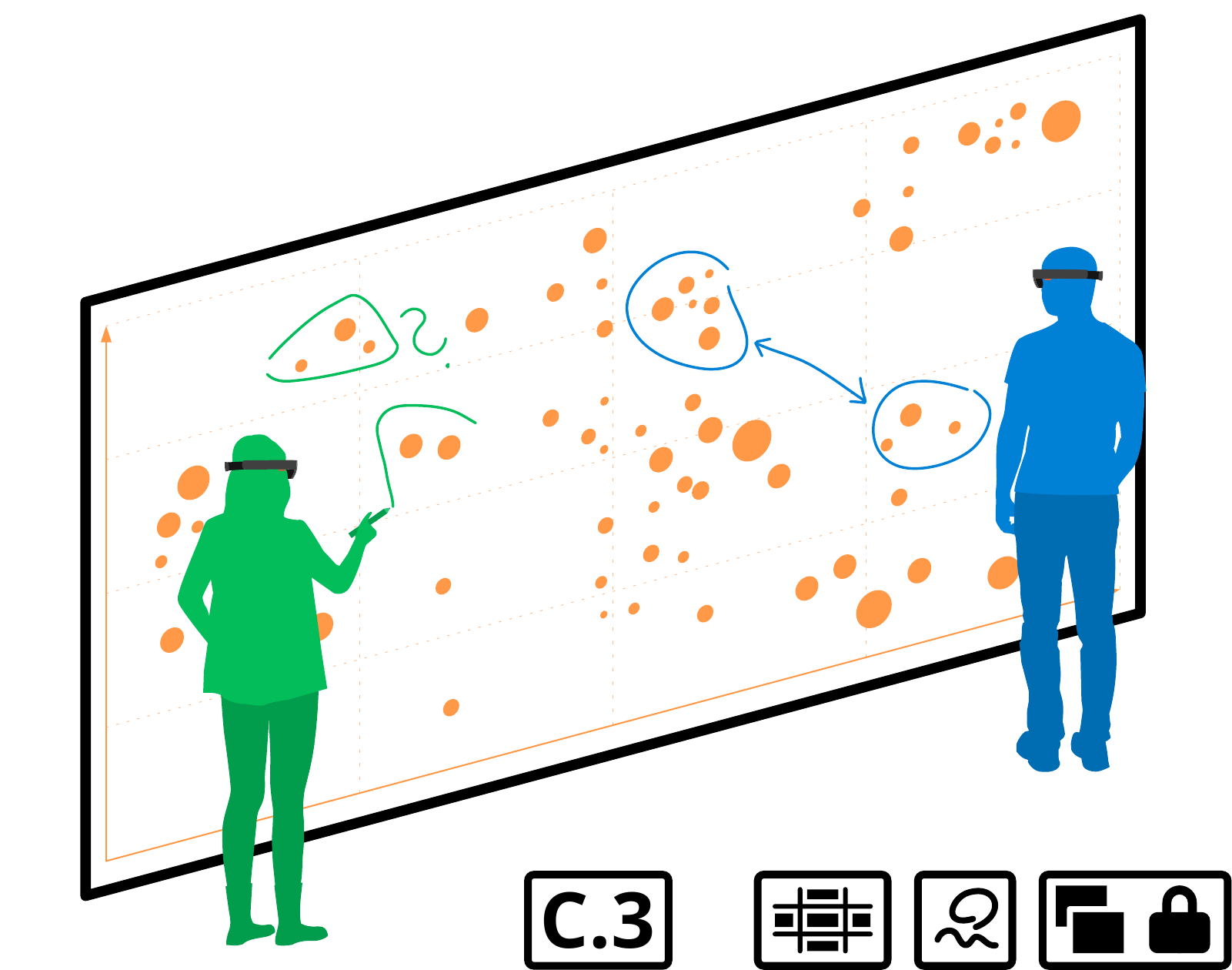}
		\caption{Personal \vtAnnotate created by pen input in green for one analyst and blue for the other.}
		\label{fig:vt:annotate}
	\end{minipage}%
\vspace{-1.0em}
\end{figure*}

\subsubsection{Curved AR Screen}\label{sec:vtCurved}
Similar to \vtHinged, this technique addresses the perceptual problem of parts of the display being in the analyst's peripheral vision (\pPercep).
However, instead of individual visualizations, this technique aims to provide an overview of and awareness for the entire screen.
Curved displays improve the perception of peripheral content, as shown by Shupp et al.~\cite{Shupp2009}.
Therefore, we propose to simulate such a curved display in AR.
The screen, with the exception of an area of configurable size directly in front of the individual \cl{analyst}, is transferred into AR space onto two curvatures, which extend to the left and right of an analyst (see Fig.~\ref{fig:vt:curved} and \ref{fig:teaser}d).
The curvatures are influenced by an analyst's position.
The further away they are from a certain display edge, the stronger the respective side's curvature becomes.
Accordingly, the curvature flattens and even vanishes completely as an analyst approaches the corresponding display edge.
The distance of an analyst to the display also reduces the curvature as they gradually move from a detailed view to an overview position.
\cl{\vtCurved supports multiple users by providing a personalized view based on each user's position and allows to see content that might otherwise be occluded by other users (\pMultiUser).}
\vtCurved affects the whole central area of the display but may also extend into the left and right areas adjacent to the display.
However, it combines well with other techniques, because content created by other techniques can be transferred onto the curvature as well.
\vtCurved affects all parts of a visualization, in fact the whole display.
As the whole screen is curved, the size and number of visualizations on the display is not relevant and the technique works well even for a single large visualization.

\subsubsection{AR Brushing and Linking}\label{sec:vtBrushLink}
When an analyst selects one or more items, connected items in other visualizations are often highlighted by Brushing and Linking.
However, they can be hard to perceive due to impaired color perception in the peripheral field of view~\cite{Andrews2011} and an analyst's ability to track changes over long distances~\cite{Endert2011} (see \pPercep).
To address this problem, three-dimensional AR links, like Bézier curves, can be displayed whenever an analyst selects data marks (see \autoref{fig:teaser}a,b, \autoref{fig:vt:brushLink} and \autoref{fig:useCase_collage}d).
This provides a clear visual link that extends from the original selection to the linked data and is therefore more easily perceptible than highlights.
These links can encode additional values, for example using the color and height~\cite{Yang:2019:FlowMaps} or the shape of the curve~\cite{Bueschel:2019:LinkStyles}.
Analysts can also filter the links, e.g., to display only links of certain attributes, thus further enhancing the analysis process and contributing to \pComplexFull.
With the number of selections or their individual size growing, the amount of links can lead to visual clutter.
To avoid this, we suggest displaying links on demand, e.g., showing them while an analyst touches a data mark, or displaying them only for a short amount of time to establish the necessary awareness.
Alternatively, the links can be clustered or bundled~\cite{Prouzeau:2019:LinkRouting,Edge2018,Lhuillier2017} but this makes it difficult to follow individual links.
In general, AR links can be used to augment \cl{data marks, links, and selections}.
They reside in the central zone of the component they augment and the central zone of the display itself because they can cross several visualizations.
\cl{Links are created individually for each person based on their current selection and therefore do not disturb others (\pMultiUser).}
As their main purpose is to show connections to other visualizations, AR links work best in a multiple coordinated view environment.
However, they might be used in single visualizations as well to show additional relations between individual data marks.
Additionally, AR links can indicate completely different types of connections, such as the dynamic information flow between visualizations.

\subsection{\pComplexName}
We now present four exemplary techniques that are primarily aimed at making dense and complex data easier to experience and explore (\pComplex).

\subsubsection{Extended Axis Views}\label{sec:vtExtendedViews}
The huge amount of data presented on a large display can be overwhelming and make it challenging to get an overview or identify patterns at first (see \pComplex).
Aggregations are an effective way to address this problem~\cite{Yost2007}.
They can be shown for specific axes, i.e., data dimensions, of a visualization, for instance by providing a histogram of the distribution of values along an axis.
In order to use the screen space exclusively for the core visualizations, we suggest moving the \vtExtendedViews into AR space (see \autoref{fig:teaser}a,b, \autoref{fig:vt:extended}, and \autoref{fig:useCase_collage}a).
\vtExtendedViews are situated directly on the axis of a visualization and expand it to the corresponding adjacent spatial zone, e.g., \cl{the bottom spatial zone} for the bottom horizontal axis.
This creates a strong spatial alignment between the axis and the extended view and thus allows to easily see which data is aggregated.
In case of wall-sized displays, the aggregation can be visualized directly on the floor in front of the wall and therefore integrates nicely with the room geometry.
When a multitude of adjacent visualizations results in constrained space, the \vtExtendedViews can be folded as described in~\autoref{sec:ds:partsOfVis}.

\cl{Besides aggregations, this technique can also be used to show additional, personalized (\pMultiUser), axis-related data for a visualization.}
For example, a 2D scatter plot can be extended to a scatter plot matrix with three dimensions.
The additional dimension is extended directly at the axis of the original visualization so that a cuboid is created (cf. ScatterDice~\cite{Elmqvist2008}).
This results in a strong spatial relation between the individual parts of the matrix.
The cuboid could be rotated with a touch gesture, so that one of the parts formerly in AR is now shown on the display in better detail and can be interacted with using touch input.

\subsubsection{AR Visualization Layers}\label{sec:vtLayer}
Dense visualizations may result in overplotting~\cite{Sarikaya2018} and an overall unreadable visualization (see \pComplex).
The increased space provided by large displays can reduce the problem of overlapping data marks but leaves remaining issues, such as edge crossings in node-link diagrams~\cite{Andrews2011}.
To tackle these issues, we propose to transfer elements of the original visualization into AR onto multiple layers or show additional data on AR visualization layers parallel to the screen \refx{fig:vt:layers}{}.
On the one hand, the \cl{layers can be used to group data marks semantically}, like filtering by specific attributes.
In case of node-link diagrams, overlapping edges also benefit from AR by resolving the ambiguity among the edges.
Edge crossings are further addressed by moving a single node, a group of nodes, or semantic clusters onto layers in front of the display.
Thereby, the connectivity of the transferred nodes to the remaining graph is evident since the connecting edges are shown between the layers.
On the other hand, additional visualizations of the same type can be placed in layers in front of or behind the display in AR allowing for superimposed comparison~\cite{Gleicher2011} and thus addressing \pComplex.
For example, if a bar chart visualizes the number of movies released in each year in a specific genre, the development for other genres can be displayed as \vtLayer for comparison.
By scrolling through the layers using a designated touch gesture, analysts can move a certain layer onto the display to interact with it and see it in more detail.
As a potential problem, the readability of the original visualization is affected by the superimposed AR layers.
We propose two solutions to overcome this problem:
First, the \vtLayer are only displayed on-demand to provide a quick summary of similarities among the data.
Second, the transparency of the AR visualizations can be reduced in favor of improved readability of the display.
Both variants can be combined, i.e., showing the AR visualizations in high transparency and switching to full opacity on-demand.
The layers themselves are either directly on, in front of, or behind the display or a specific visualization, taking up the central spatial zone.
\vtLayer affect data marks, links, \cl{and selections}, and may even augment the whole visualization.
While the additional AR visualizations are only visible to individual analysts, scrolling through the \vtLayer may \cl{obstruct other analysts that intend to work with previous layers (\pMultiUser)}.

\subsubsection{Magic AR Lenses}\label{sec:vtMagicLenses}
Complex data sets benefit from using specialized tools to facilitate analysis and exploration.
These tools can also be transferred to AR space.
In the following, this is illustrated by the example of magic lenses, which have been successfully applied to the field of data visualization in the past~\cite{Tominski:2017:MagicLenses}.
\cl{Magic lenses can apply arbitrary filters to the underlying data set, e.g., to reduce its complexity (\pComplex), which makes them a useful, personalizable tool for data analysis.
However, these tools take up a lot of space on the visualization and can thus disturb other analysts in multi-user scenarios during exploration.
They may also extend the current data by private information that analysts do not want to share.}
By transferring a lens into AR space, it becomes invisible to other analysts \refx{fig:vt:magicLenses}{}.
Not only are they no longer hindered by the lens, but the lens becomes a personal, user-specific tool that can be freely configured according to an analyst's needs (\pMultiUser).
Controlling the lens can still be achieved via the touch interface of the display.
If the spatial alignment between the display and the AR device is synchronized, the analyst can interact with the AR object as if it were still on the screen.
In addition, the filters assigned to a lens can now use the third dimension added by AR which enables an entirely new class of filters, such as resolving the overlapping of nodes (similar to \vtLayer).
\vtMagicLenses mostly affect data marks but may, depending on the applied filters, affect other parts of the visualization as well.
The lenses themselves reside directly on and are aligned with the display or individual visualization but filters may expand the lenses into the space in front or even behind the screen.
Magic Lenses are suitable for single as well as multiple visualizations on the display, although an individual lens is usually confined to only a single visualization, e.g., a map.

\subsubsection{AR Annotations}\label{sec:vtAnnotate}
While analyzing data, it is often helpful for analysts to create notes, sketches, and drawings directly on the visualizations to better capture insights\cite{Walny2012,Walny2018}.
\cl{However, the notes of one analyst can be irrelevant and even disturbing for others} and influence their own analysis process, which is an essential part of \pMultiUser.
\cl{At the same time, analysts may want to keep their notes private and not immediately share them with others.}
Therefore, we suggest moving the annotations to AR space.
Analysts create their annotations writing on the display, for example using a digital pen.
However, these annotations are created solely in AR space (see \autoref{fig:vt:annotate} and \autoref{fig:useCase_collage}b).
Provided the quality of the AR content is sufficiently good, there is no difference between the presentation of annotations on the display from the perspective of an individual analyst.
Furthermore, analysts should be enabled to share their annotations with others using additional touch interaction techniques on the display to foster collaboration.


\subsection{Combining Techniques}

Many of the previously presented techniques can be adapted or even combined according to the requirements of a specific use case.
Techniques that affect different parts of a visualization can be \cl{used together} without conflict.
The same applies to techniques that affect different spatial zones.
In addition, techniques can be adapted to be compatible with each other to create a mutually beneficial combination.
For example, \vtMagicLenses can show \vtEmbedded for all data marks in the lens area allowing to select whole regions instead of individual data marks.
\vtMagicLenses can also be combined with \vtLayer to show layers only for a specific area rather than for the whole visualization, which can be advantageous for large visualizations.
\vtBrushLink can be adapted to not only link visualizations on the display but also visualizations residing in AR like \vtHinged, \vtCurved, and \vtExtendedViews.
Furthermore, principles like \vtHinged can be applied to techniques such as \vtExtendedViews and even \vtMagicLenses.
Therefore, those visualizations would face the analyst as well and ensure better readability over far distances.

%% file: content/prototype.tex
\section{AR Visualization Framework and Prototype}\label{sec:prototype}
In order to \cl{execute and examine} our techniques, we first developed a visualization framework that enables traditional visualizations on a display as well as their free placement in the 3D AR space and a combination of the two.
Furthermore, it provides extensive capabilities for manipulation and interaction with the visualizations.
This was used as a foundation for the prototype combining a display and AR HMDs and implementing the techniques presented in the previous section.

\parainline{Technical Setup}
As a large display we used a Microsoft \emph{Surface Hub} (84", 3840x2160 pixel, 1.872x1.053m, touch and pen enabled) and our own tiled display wall (7680x3240 pixel, 4.86x2.06m, touch, pen, and tangible enabled).
For Augmented Reality, we provide a MS \emph{HoloLens~1} head-mounted display for each user.
We used the \emph{Unity} 3D engine and C\# to implement all parts of the prototype and framework.
The application on the display and those on the HMDs are stand-alone and initially independent.
Since the impression of a single, seamless application is of great importance for the user experience, we developed a custom client server solution using TCP and Open Sound Control (OSC) to synchronize them especially regarding the touch interaction of the users.

\begin{figure}[tb]
	\includegraphics[width=\columnwidth]{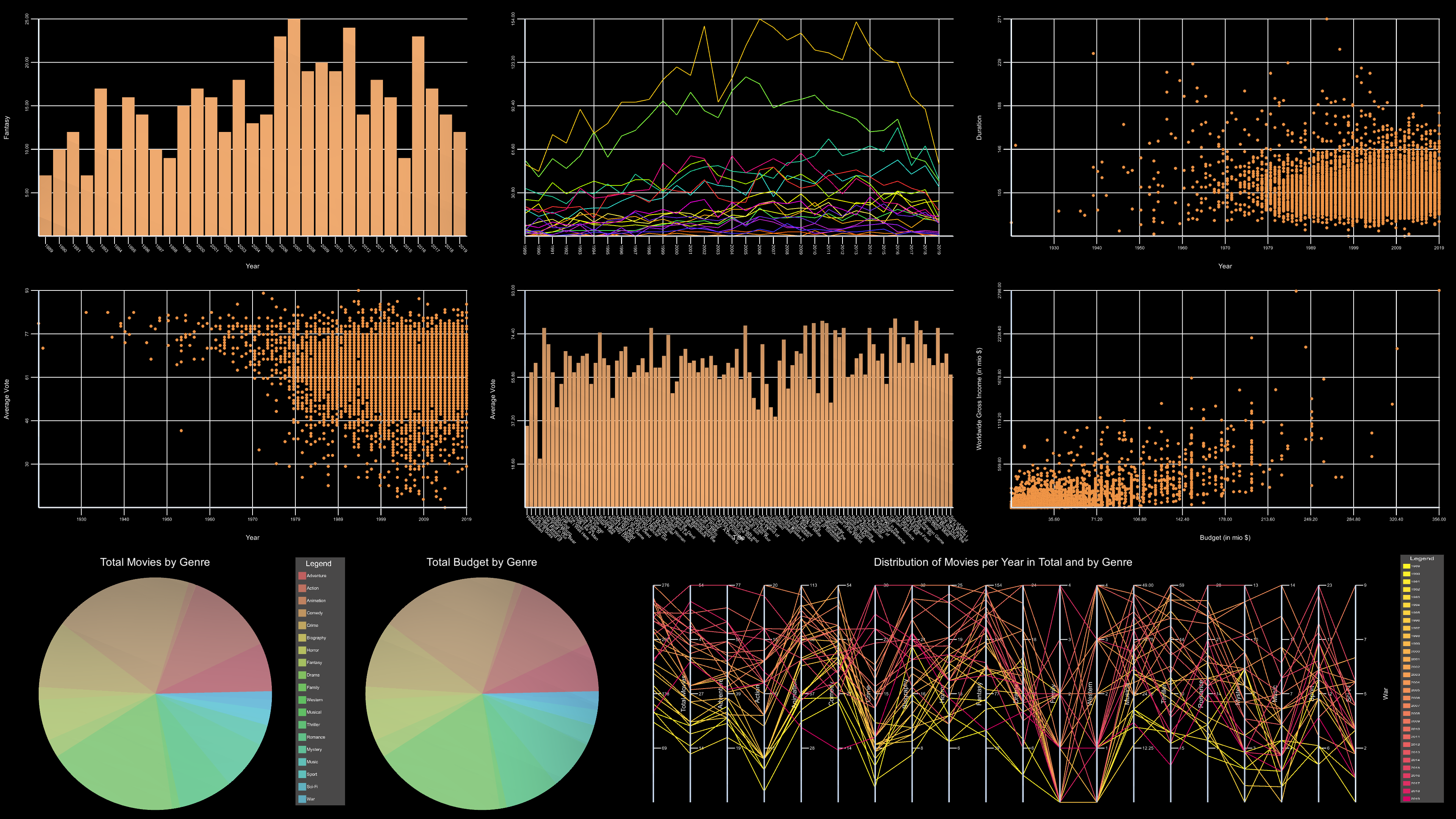}
	\caption{Screenshot of the \emph{Unity} application running on the display, showcasing the capabilities of our visualization framework.
	Several visualizations of a movie data set are shown that serve as the foundation for our use case.}
	\label{fig:screenshot}
	\vspace*{-1.5em}
\end{figure}

\parainline{u2vis: A Universal Unity Visualization Framework}
Although many frameworks for information visualization exist, they are mostly limited to traditional 2D visualizations.
\cl{We investigated whether existing frameworks for \emph{Immersive Analytics} are suitable for our goals.
IATK\footnote{\small{\url{https://github.com/MaximeCordeil/IATK}}} looked promising but we found it difficult to integrate the interaction capabilities, especially regarding touch, that were essential for our prototype.
DXR\footnote{\small{\url{https://sites.google.com/view/dxr-vis}}} only supports \emph{Unity 2017} and seems to not be maintained anymore.}
Therefore, we implemented our own, data-driven framework for interactive information visualization called \emph{u2vis}\footnote{Sources available at \url{https://github.com/imldresden/u2vis}}.
It provides a joint data basis for all visualizations, separates data and its presentation, and allows for the manipulation of visualizations by touch and other interaction modalities.
Our framework currently supports 2D and 3D bar charts, scatter plots, and line charts, as well as parallel coordinates and pie charts.
While \autoref{fig:screenshot} only depicts the 2D projection shown on the display, all visualizations can be freely positioned in 3D as well.
Our framework is written with an explicit focus on easy extensibility, making it simple and straightforward to add new visualization types.
All visualizations are composed by combining several components in the \emph{Unity} editor by drag and drop and configured with the help of a graphical interface.
The loose coupling between these components makes it possible to use only needed parts and to easily replace components with alternatives tailored to the current use case.

\parainline{Prototype}
Based on this foundation, we implemented the techniques presented in~\autoref{sec:techniques} as a prototype \refx{fig:teaser}{}.
Visualizations can be placed in any arrangement on the display and each user is able to access their desired techniques independent of other users.
Since the individual techniques build upon the common foundation of our visualization framework, combining them with each other is easy.
To ensure the spatial alignment of the AR content and the screen, the AR content has a shared anchor on the lower left corner of the display.
All objects in AR are positioned relative to the anchor.
For interaction, we focused on touch input to interact with the visualizations and our techniques and use the digital pen to create personal annotations.
When a user touches the display, the corresponding action, e.g., creating a selection, is computed on the display and the result, e.g., highlighting a specific data object, is transmitted to the HMDs.
Within our framework, each visualization consists of a display part, which can function on its own, and the HMD part, that reacts to events from the display's counterpart.

%% file: content/useCase.tex
\section{Use Case: Exploring a Movie Data Set}\label{sec:useCase}

\cl{
We examine our concepts as well as our implemented techniques in the form of a cognitive walkthrough based on well-established analysis tasks~\cite{Amar2005} and a specific use case scenario~\cite{Ledo2018}, that we implemented in our prototype.
We visualize a movie data set\footnote{available on Kaggle {\tiny (\url{https://www.kaggle.com/stefanoleone992/imdb-extensive-dataset})}} consisting of multiple coordinated views \refx{fig:screenshot}{}.
In the following, we describe how \userX and \userY solve analysis and exploration tasks utilizing our techniques.

\begin{figure}
	\centering
	\includegraphics[width=0.97\columnwidth]{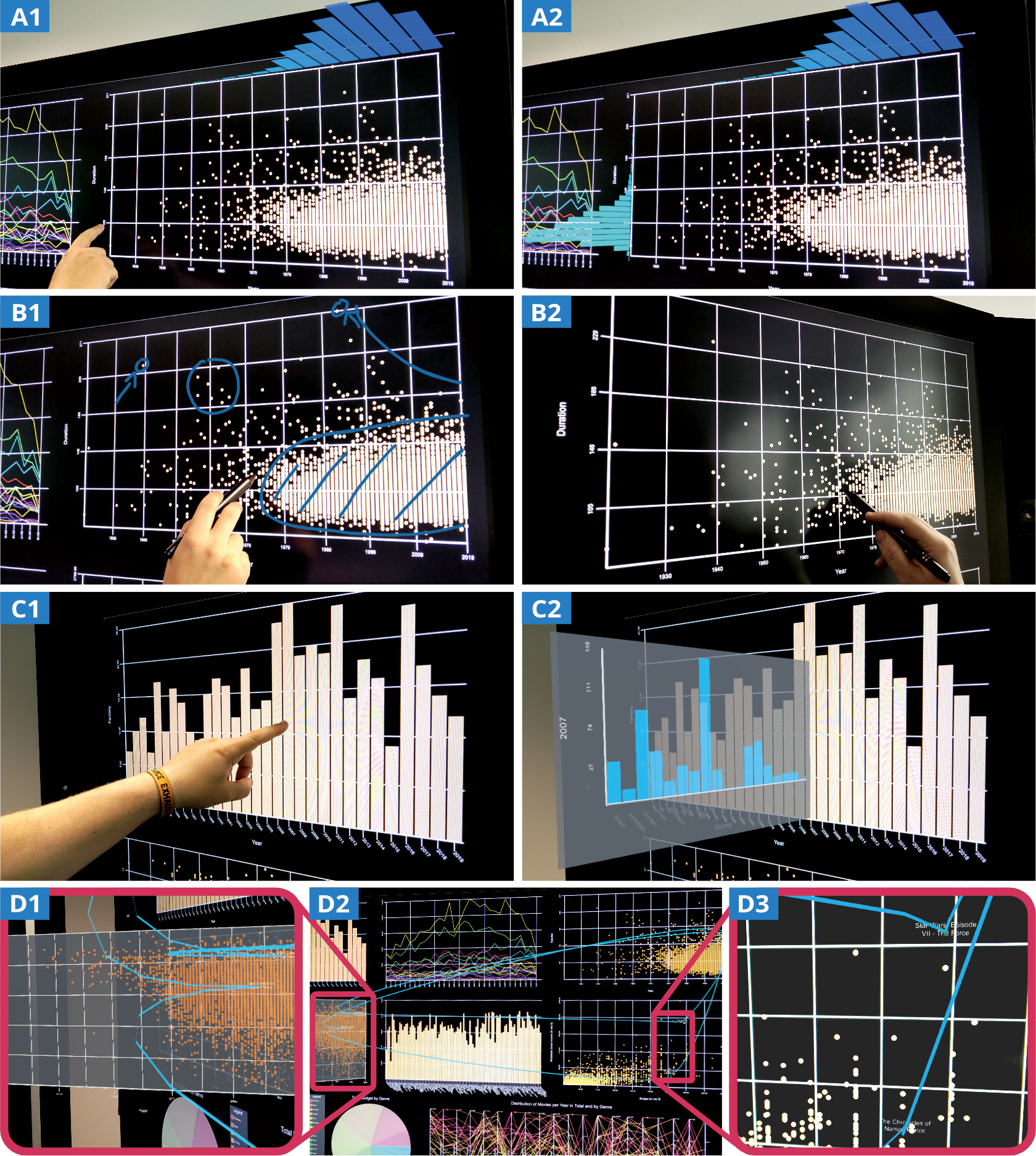}
	\caption{\cl{Illustrating our use case: (a1) Selecting \vtExtendedViews that (a2) show an aggregation of movie duration. (b1) One analyst annotating a scatter plot, (b2) without the sketches being visible to others. (c1) Selecting the year 2007 to show (c2) \vtEmbedded of the amount of movies released in each genre for this year. (d2) \vtBrushLink between visualizations, (d3) a selection of two movies in a scatter plot visualizing budget and worldwide gross income, (d1) \vtHinged showing the average votes for those movies}.}
	\label{fig:useCase_collage}
	\vspace{-1.5em}
\end{figure}

\vspace{0.8em}\noindent
\userX and \userY decide to investigate the typical duration of movies and how it changed over the years.
They approach a scatter plot showing the release date from 1920 to 2019 on the x axis and the duration of a movie on the y axis.
To better identify clusters, both choose our \emphref{sec:vtExtendedViews} technique to show a histogram at the y axis \refx{fig:useCase_collage}{a}.
Because the histogram is displayed in AR, no other visualizations needs to be replaced, moved, or overlaid.
With the help of the histogram \userY detects that the typical movie duration is between 90 and 108 minutes.
The scatter plot itself reveals that movies nowadays tend to have similar durations but with fewer outliers than in the 20th century.
To accentuate this development, \userY sketches with his digital pen to create personal \emphref{sec:vtAnnotate} \refx{fig:useCase_collage}{b}.

The two analysts now focus on separate tasks, so they decide to start working independently.
\userX is further interested in the development of movies over time, especially in fantasy movies, whereas \userY wants to gain knowledge about specific movies.
Therefore, \userX moves towards a bar chart showing the number of fantasy movies over time. 
A hold gesture on a data mark, i.e., a specific year, reveals another bar chart orthogonal to the display, using the \emphref{sec:vtEmbedded} technique, which shows the number of movies in all genres for that particular year \refx{fig:useCase_collage}{c}.
While the most fantasy movies (25) were produced in 2007, \userX finds out that this is still vastly below the number of produced dramas in that year (148).
Through \vtEmbedded, \userX can access this information on demand directly within the context of an existing visualization, avoiding focus switches and saving valuable screen space.
%

Meanwhile, \userY looks at the scatter plot showing budget and worldwide gross income and selects two movies with similar budget but a huge difference in income, which are \emph{Star Wars: Episode VII - The Force Awaken} and \emph{The Chronicles of Narnia: Prince Caspian}.
With our \emphref{sec:vtBrushLink} technique, the selected movies are highlighted in all other relevant visualizations with 3D links in AR \refx{fig:useCase_collage}{d}.
This, together with \emphref{sec:vtHinged} (or alternatively the \emphref{sec:vtCurved}) guides his awareness to the relevant visualizations.
Thus, he does not have to move to the other side of the display to obtain an overview and an approximate position of the selected data. Retrieving exact values, however, requires him to approach the visualization.
One of the \vtHinged shows a scatter plot of release date in comparison to average user vote.
Based on the selection and the accompanying links, it is easy to spot the desired movies in this chart and to derive that \textit{Star Wars} has an average rating of $7.9$ and \textit{Narnia} of $6.5$.
Since only \userY sees the selection, the links, and the curved or hinged parts, \userX can keep exploring the data without visual clutter or disturbance.

\vspace{0.8em}\noindent
This cognitive walkthrough shows how our proposed setup supports common analysis tasks and highlights the potential benefits compared to traditional setups, such as saving valuable screen space, integrating additional information directly in place, and not obstructing other analysts.
Based on this use case and our own experiences, we are confident that our approach has high potential in assisting data analysis involving multiple users.
However, this claim must be evaluated in a user study with specific analysis tasks that need to be solved using our setup in comparison to conventional multiple coordinated views systems (cf.~\cite{Langner2019}).
}

%% file: content/discussion.tex
\section{Discussion}\label{sec:discussion}
To summarize our experiences based on the prototype and the cognitive walkthrough from the use case, we discuss the perception of AR content situated on the display as well as the technical limitations of our setup and state-of-the-art AR in general.
Furthermore, we consider collaborative work and possible future applications.

\parainline{Perception of AR Content and Display}
An obvious question when extending or overlaying a display with AR content is how well each medium can be perceived and if they influence each other.
In the case of AR content, this largely depends on the technology.
The \textit{HoloLens~1} we used creates a relatively opaque image with little bleed-through of the background.
However, resolution and field of view are limited, so displaying very small or detailed content is problematic.
This is especially true for very small text.
We recommend to show such content only on the display as its high resolution predestines it for these fine details.
Furthermore, it should be noted that the perception of colors in AR is inferior to conventional displays~\cite{Whitlock2020}.
Therefore, visual attributes, such as shape or size, should be preferably used since they are well perceptible in AR.
Otherwise, displaying AR content superimposed on the display is usually unproblematic with respect to perception.

Regarding the display itself, a possible issue is that AR content makes it difficult to see the visualizations \textit{on} the display.
AR content should therefore not be placed arbitrarily but always in relation to objects on the display, so that they support and extend them.
Nonetheless, switching between AR and display content may still lead to focus and attention issues.
Thus, we recommend to carefully examine which content to transfer to AR and whether its benefits outweigh its costs.
However, AR can also reduce focus switches, for instance through \vtEmbedded.
Furthermore, we suggest the use of on-demand techniques to reduce the number of objects displayed simultaneously.
In general, Augmented Reality technology still has lots of room for improvement and we expect it to further advance over the coming years.
We believe that the visual quality of AR displays might become similar to the quality of high-resolution displays.

One question that arises in this context is: Why is an interactive display still needed at all if the quality of an AR HMD is sufficiently good?
First, the display has a haptic affordance not readily available in AR and analysts can interact with data using natural touch and pen input.
Second, the display is visible to everyone without requiring additional technology and can be used as a shared space for data exploration.
Third, the display offers a physical point of reference which can be used for the orientation and alignment of AR content.

\parainline{Combining Environments for Data Analysis}
Using different interactive environments has great potential for data analysis which was demonstrated by previous work on large display environments~\cite{Andrews2010,Langner2019}, physical visualizations~\cite{Dragicevic2020}, and immersive environments~\cite{Book:ImAn2018}.
We believe that combining different environments can leverage their advantages and mitigate their disadvantages at the same time.
In this case, interaction constitutes a special challenge due to switches between interaction paradigms.
\cl{Similar to Butcher et al.~\cite{Butscher2018}, our concepts address the missing haptic feedback of mid-air interaction commonly used in AR by using touch interaction to control AR.}
However, possibly having to reach through AR objects may nevertheless seem strange to users due to obscuring their hands.
Furthermore, the optimal balance between content on the display and content situated in AR is an interesting question that calls for further study. 
In this context, we have proposed an initial approach through our concepts.
Nevertheless, we agree with Bach et al.~\cite{Bach2018a} on not advocating for a complete substitution of existing environments and techniques.
We consider our approach to be an interesting addition to other established strategies to address the known issues \refx{sec:challenges}{} of large display environments and to make them an even more powerful and sophisticated tool for data analysts, especially for multi-user scenarios.

\parainline{Alignment of AR Content in Relation to the Display}
One of our assumptions in this work is that the alignment of AR content with the display has a positive effect on analysts' performance.
However, it is unclear how this impacts the user's experience and it should be further examined in comparison to alternative placement options.
As mentioned in~\autoref{sec:ds:spatialrelation}, a precise spatial alignment of the AR content with the content shown on display is essential for many of the presented techniques.
This places great demands on the tracking technology used to ensure that, for instance, AR objects bound to specific data marks in a dense visualization are not mistakenly associated with adjacent data marks.
With the \textit{HoloLens~1} we used for our prototype, limited accuracy and drift are major issues that get worse with increasing display sizes.
Feature-based tracking, as used by the \textit{HoloLens}, reaches its limits when users are extremely close to a large or wall-sized display as only few features remain that can be tracked.
Furthermore, the alignment of objects is not stable regarding rotation and translation.
As a consequence, objects constantly shift their position as analysts move around in front of the display.
We expect these problems to improve and eventually be solved with subsequent technologies.

\parainline{Collaboration Between Multiple Analysts} \label{sec:dis:collaboration}
\cl{
By using AR HDMs, our setup currently has a technology-caused intrinsic personal view for all analysts.
A system that would be focused on fostering communication and collaboration requires dedicated techniques for sharing content between analysts.
However, simply sharing all content between everyone wastes much of the inherent potential of the system, as personalized views can vary greatly from person to person or analysts may want to keep their annotations private.
The resulting interesting design question is which content should be visible to everyone or only to certain analysts and which content should remain private.
This, however, is beyond the scope of this work and left for future investigation.
In addition, we refer to the large body of work in the field of computer-supported cooperative work (CSCW) for general techniques to support collaboration.
}

\cl{
\parainline{Connecting Multiple Visualizations with AR}
While AR can be used to better illustrate relationships between several visualizations (see \vtBrushLink), there is potential in using AR to visualize the connection in more detail.
AR could be used to create temporary, user-specific views derived from the connection between existing visualizations.
For example, an AR view can highlight the difference between two visualizations and aid in comparison tasks~\cite{Gleicher2018, Gleicher2011}.
We think there is merit in investigating the types of connections that can visualized and how to best represent them using AR.
}

\parainline{Further Potential Utilization of AR}
In the context of this work, we have so far talked about the direct extension of the display with AR content.
However, the use of AR content that is independent of the display but is instead based on the spatial position of the user can be useful as well.
One possible approach is to move away from simply extending content on the display and to focus on the AR visualizations.
In this case, the display is still valuable for naturally interacting with the AR content. 
For example, a 3D AR visualization, such as a height map, could be positioned on the display.
Specialized tools, such as slicing planes, can then be manipulated using touch interaction to facilitate a convenient way for the exploration of 3D visualizations.
Although our chosen setup has a particular set of affordances and requirements, our techniques can serve as an inspiration for further adaption.
For example, they can be extended to horizontal displays and even mobile devices as well.
They are not limited to information visualization but can be applied to data and scientific visualization as well.

%% file: content/conclusion.tex
\section{Conclusion}\label{sec:conclusion}

We presented the combination of a large interactive display with head-mounted, personal Augmented Reality for information visualization.
Our contribution consists of a design space, techniques for extending visualizations with AR, and the data-driven AR visualization framework \emph{u2vis} as well as the accompanying prototype we developed.
\cl{We further investigated our approach with a cognitive walkthrough of a use case scenario, which indicates that our techniques can help to reduce perceptional issues in large displays for data exploration and to facilitate the analysis of dense data sets.
However, further and more extensive evaluation is required to validate these indications.}
We are therefore convinced that the extension of large displays with AR has great potential for information and data analysis.
We hope that our work inspires the exploration of this new exciting class of visualization applications, which combine interactive displays with Augmented Reality.